\RequirePackage{lineno}
\documentclass[twocolumn,aps,prd,floatfix,superscriptaddress]{revtex4}
\usepackage{latexsym, amsmath, amssymb,amsthm,amsopn,amsfonts, graphicx, epstopdf, multirow}
\usepackage{color}
\usepackage{amssymb}
\usepackage{amsmath}
\usepackage{natbib}
\usepackage{textcomp}
\usepackage{placeins}
\usepackage[dvipsnames]{xcolor}
\usepackage{adjustbox}
\usepackage{cancel}
\usepackage{multirow}
\usepackage{orcidlink}
\usepackage{hyperref}

\newcommand\bk{BICEP/\textit{Keck }}
\renewcommand\deg{^\circ}
\newcommand{\Cl}{\mathcal{C}_\ell}
\newcommand{\Cb}{\mathcal{C}_b}
\newcommand{\oCb}{\mathcal{\hat C}_b}
\newcommand{\rCb}{\mathcal{C'}_b}
\newcommand{\Ccov}{\mathrm{\mathbf{C}}}

\newcommand{\cov}{\text{Cov}}
\newcommand{\tozero}[1]{\color{red}{\bcancel{\color{black}{#1}}}\color{black}}

\definecolor{matblue}{rgb}{0 0.447 0.741}
\definecolor{matred}{rgb}{0.85 0.3250 0.098}

\hypersetup{linktoc=all}
\hypersetup{colorlinks = true, allcolors = blue}

\begin{document}

\title{BICEP/\textit{Keck} XVIII: Measurement of BICEP3 polarization angles \\ and consequences for constraining cosmic birefringence and inflation}

\author{BICEP/\textit{Keck} Collaboration:~P.~A.~R.~Ade}
\affiliation{School of Physics and Astronomy, Cardiff University, Cardiff, CF24 3AA, United Kingdom}

\author{Z.~Ahmed\orcidlink{0000-0002-9957-448X}}
\affiliation{Kavli Institute for Particle Astrophysics and Cosmology, Stanford University, Stanford, CA 94305}
\affiliation{SLAC National Accelerator Laboratory, Menlo Park, CA 94025, USA}

\author{M.~Amiri\orcidlink{0000-0001-6523-9029}}
\affiliation{Department of Physics and Astronomy, University of British Columbia, Vancouver, British Columbia, V6T 1Z1, Canada}

\author{D.~Barkats\orcidlink{0000-0002-8971-1954}}
\affiliation{Center for Astrophysics $\vert$ Harvard \& Smithsonian, Harvard University, Cambridge, MA 02138, USA}

\author{R.~Basu Thakur\orcidlink{0000-0002-3351-3078}}
\affiliation{Department of Physics, California Institute of Technology, Pasadena, CA 91125, USA}

\author{C.~A.~Bischoff\orcidlink{0000-0001-9185-6514}}
\affiliation{Department of Physics, University of Cincinnati, Cincinnati, OH 45221, USA}

\author{D.~Beck\orcidlink{0000-0003-0848-2756}}
\affiliation{Department of Physics, Stanford University, Stanford, CA 94305, USA}

\author{J.~J.~Bock}
\affiliation{Department of Physics, California Institute of Technology, Pasadena, CA 91125, USA}
\affiliation{Jet Propulsion Laboratory, California Institute of Technology, Pasadena, CA 91109, USA}

\author{H.~Boenish}
\affiliation{Center for Astrophysics $\vert$ Harvard \& Smithsonian, Harvard University, Cambridge, MA 02138, USA}

\author{V.~Buza}
\affiliation{Kavli Institute for Cosmological Physics, University of Chicago, Chicago, IL 60637, USA}

\author{J.~R.~Cheshire IV\orcidlink{0000-0002-1630-7854}}
\affiliation{Department of Physics, California Institute of Technology, Pasadena, CA 91125, USA}
\affiliation{Minnesota Institute for Astrophysics, University of Minnesota, Minneapolis, MN 55455, USA}

\author{J.~Connors}
\affiliation{National Institute of Standards and Technology, Boulder, CO 80305, USA}

\author{J.~Cornelison\orcidlink{0000-0002-2088-7345}}
\email{james.a.cornelison@gmail.com}
\affiliation{Center for Astrophysics $\vert$ Harvard \& Smithsonian, Harvard University, Cambridge, MA 02138, USA}

\author{M.~Crumrine}
\affiliation{School of Physics and Astronomy, University of Minnesota, Minneapolis, MN 55455, USA}

\author{A.~J.~Cukierman}
\affiliation{Department of Physics, Stanford University, Stanford, CA 94305, USA}

\author{E.~Denison}
\affiliation{National Institute of Standards and Technology, Boulder, CO 80305, USA}

\author{L.~Duband}
\affiliation{Service des Basses Temperatures, Commissariat a l’Energie Atomique, 38054 Grenoble, France}

\author{M.~Eiben}
\affiliation{Center for Astrophysics $\vert$ Harvard \& Smithsonian, Harvard University, Cambridge, MA 02138, USA}

\author{B.~D.~Elwood\orcidlink{0000-0003-4117-6822}}
\affiliation{Department of Physics, Harvard University, Cambridge, MA 02138, USA}
\affiliation{Center for Astrophysics $\vert$ Harvard \& Smithsonian, Harvard University, Cambridge, MA 02138, USA}

\author{S.~Fatigoni\orcidlink{0000-0002-3790-7314}}
\affiliation{Department of Physics, California Institute of Technology, Pasadena, CA 91125, USA}

\author{J.~P.~Filippini\orcidlink{0000-0001-8217-6832}}
\affiliation{Department of Physics, University of Illinois at Urbana-Champaign, Urbana, IL 61801, USA}
\affiliation{Department of Astronomy, University of Illinois at Urbana-Champaign, Urbana, IL 61801, USA}

\author{A.~Fortes}
\affiliation{Department of Physics, Stanford University, Stanford, CA 94305, USA}

\author{M.~Gao}
\affiliation{Department of Physics, California Institute of Technology, Pasadena, CA 91125, USA}

\author{C.~Giannakopoulos}
\affiliation{Department of Physics, University of Cincinnati, Cincinnati, OH 45221, USA}

\author{N.~Goeckner-Wald}
\affiliation{Department of Physics, Stanford University, Stanford, CA 94305, USA}

\author{D.~C.~Goldfinger\orcidlink{0000-0001-5268-8423}}
\affiliation{Department of Physics, Stanford University, Stanford, CA 94305, USA}

\author{J.~A.~Grayson}
\affiliation{Department of Physics, Stanford University, Stanford, CA 94305, USA}

\author{A. Greathouse}
\affiliation{School of Physics and Astronomy, University of Minnesota, Minneapolis, MN 55455, USA}

\author{P.~K.~Grimes\orcidlink{0000-0001-9292-6297}}
\affiliation{Center for Astrophysics $\vert$ Harvard \& Smithsonian, Harvard University, Cambridge, MA 02138, USA}

\author{G.~Hall}
\affiliation{School of Physics and Astronomy, University of Minnesota, Minneapolis, MN 55455, USA}
\affiliation{Department of Physics, Stanford University, Stanford, CA 94305, USA}

\author{G.~Halal\orcidlink{0000-0003-2221-3018}}
\affiliation{Department of Physics, Stanford University, Stanford, CA 94305, USA}

\author{M.~Halpern}
\affiliation{Department of Physics and Astronomy, University of British Columbia, Vancouver, British Columbia, V6T 1Z1, Canada}

\author{E.~Hand}
\affiliation{Department of Physics, University of Cincinnati, Cincinnati, OH 45221, USA}

\author{S.~A.~Harrison}
\affiliation{Center for Astrophysics $\vert$ Harvard \& Smithsonian, Harvard University, Cambridge, MA 02138, USA}

\author{S.~Henderson}
\affiliation{Kavli Institute for Particle Astrophysics and Cosmology, Stanford University, Stanford, CA 94305}
\affiliation{SLAC National Accelerator Laboratory, Menlo Park, CA 94025, USA}

\author{J.~Hubmayr}
\affiliation{National Institute of Standards and Technology, Boulder, CO 80305, USA}

\author{H.~Hui\orcidlink{0000-0001-5812-1903}}
\affiliation{Department of Physics, California Institute of Technology, Pasadena, CA 91125, USA}

\author{K.~D.~Irwin}
\affiliation{Department of Physics, Stanford University, Stanford, CA 94305, USA}

\author{J.~H.~Kang\orcidlink{0000-0002-3470-2954}}
\affiliation{Department of Physics, California Institute of Technology, Pasadena, CA 91125, USA}

\author{K.~S.~Karkare\orcidlink{0000-0002-5215-6993}}
\affiliation{Kavli Institute for Particle Astrophysics and Cosmology, Stanford University, Stanford, CA 94305}
\affiliation{SLAC National Accelerator Laboratory, Menlo Park, CA 94025, USA}

\author{S.~Kefeli}
\affiliation{Department of Physics, California Institute of Technology, Pasadena, CA 91125, USA}

\author{J.~M.~Kovac\orcidlink{0009-0003-5432-7180}}
\affiliation{Department of Physics, Harvard University, Cambridge, MA 02138, USA}
\affiliation{Center for Astrophysics $\vert$ Harvard \& Smithsonian, Harvard University, Cambridge, MA 02138, USA}

\author{C.~Kuo}
\affiliation{Department of Physics, Stanford University, Stanford, CA 94305, USA}

\author{K.~Lau\orcidlink{0000-0002-6445-2407}}
\affiliation{Department of Physics, California Institute of Technology, Pasadena, CA 91125, USA}

\author{M.~Lautzenhiser}
\affiliation{Department of Physics, University of Cincinnati, Cincinnati, OH 45221, USA}

\author{A.~Lennox}
\affiliation{Department of Astronomy, University of Illinois at Urbana-Champaign, Urbana, IL 61801, USA}

\author{T.~Liu\orcidlink{0000-0001-5677-5188}}
\affiliation{Department of Physics, Stanford University, Stanford, CA 94305, USA}

\author{K.~G.~Megerian}
\affiliation{Jet Propulsion Laboratory, California Institute of Technology, Pasadena, CA 91109, USA}

\author{L.~Minutolo}
\affiliation{Department of Physics, California Institute of Technology, Pasadena, CA 91125, USA}

\author{L.~Moncelsi\orcidlink{0000-0002-4242-3015}}
\affiliation{Department of Physics, California Institute of Technology, Pasadena, CA 91125, USA}

\author{Y.~Nakato}
\affiliation{Department of Physics, Stanford University, Stanford, CA 94305, USA}

\author{H.~T.~Nguyen}
\affiliation{Jet Propulsion Laboratory, California Institute of Technology, Pasadena, CA 91109, USA}
\affiliation{Department of Physics, California Institute of Technology, Pasadena, CA 91125, USA}

\author{R.~O'brient}
\affiliation{Jet Propulsion Laboratory, California Institute of Technology, Pasadena, CA 91109, USA}
\affiliation{Department of Physics, California Institute of Technology, Pasadena, CA 91125, USA}

\author{A.~Patel}
\affiliation{Department of Physics, California Institute of Technology, Pasadena, CA 91125, USA}

\author{M.~A.~Petroff\orcidlink{0000-0002-4436-4215}}
\affiliation{Center for Astrophysics $\vert$ Harvard \& Smithsonian, Harvard University, Cambridge, MA 02138, USA}

\author{A.~R.~Polish\orcidlink{0000-0002-7822-6179}}
\affiliation{Department of Physics, Harvard University, Cambridge, MA 02138, USA}
\affiliation{Center for Astrophysics $\vert$ Harvard \& Smithsonian, Harvard University, Cambridge, MA 02138, USA}

\author{T.~Prouve}
\affiliation{Service des Basses Temperatures, Commissariat a l’Energie Atomique, 38054 Grenoble, France}

\author{C.~Pryke\orcidlink{0000-0003-3983-6668}}
\affiliation{School of Physics and Astronomy, University of Minnesota, Minneapolis, MN 55455, USA}

\author{C.~D.~Reintsema}
\affiliation{National Institute of Standards and Technology, Boulder, CO 80305, USA}

\author{T.~Romand}
\affiliation{Department of Physics, California Institute of Technology, Pasadena, CA 91125, USA}

\author{M.~Salatino}
\affiliation{Department of Physics, Stanford University, Stanford, CA 94305, USA}

\author{A.~Schillaci}
\affiliation{Department of Physics, California Institute of Technology, Pasadena, CA 91125, USA}

\author{B.~Schmitt}
\affiliation{Center for Astrophysics $\vert$ Harvard \& Smithsonian, Harvard University, Cambridge, MA 02138, USA}

\author{B.~Singari\orcidlink{0000-0001-7387-0881}}
\affiliation{School of Physics and Astronomy, University of Minnesota, Minneapolis, MN 55455, USA}

\author{K. Sjoberg}
\affiliation{Center for Astrophysics $\vert$ Harvard \& Smithsonian, Harvard University, Cambridge, MA 02138, USA}

\author{A.~Soliman}
\affiliation{Jet Propulsion Laboratory, California Institute of Technology, Pasadena, CA 91109, USA}
\affiliation{Department of Physics, California Institute of Technology, Pasadena, CA 91125, USA}

\author{T.~St Germaine}
\affiliation{Center for Astrophysics $\vert$ Harvard \& Smithsonian, Harvard University, Cambridge, MA 02138, USA}

\author{A.~Steiger\orcidlink{0000-0003-0260-605X}}
\affiliation{Department of Physics, California Institute of Technology, Pasadena, CA 91125, USA}

\author{B.~Steinbach}
\affiliation{Department of Physics, California Institute of Technology, Pasadena, CA 91125, USA}

\author{R.~Sudiwala}
\affiliation{School of Physics and Astronomy, Cardiff University, Cardiff, CF24 3AA, United Kingdom}

\author{K.~L.~Thompson}
\affiliation{Department of Physics, Stanford University, Stanford, CA 94305, USA}

\author{C.~Tsai}
\affiliation{Center for Astrophysics $\vert$ Harvard \& Smithsonian, Harvard University, Cambridge, MA 02138, USA}

\author{C.~Tucker\orcidlink{0000-0002-1851-3918}}
\affiliation{School of Physics and Astronomy, Cardiff University, Cardiff, CF24 3AA, United Kingdom}

\author{A.~D.~Turner}
\affiliation{Jet Propulsion Laboratory, California Institute of Technology, Pasadena, CA 91109, USA}

\author{C.~Vergès\orcidlink{0000-0002-3942-1609}}
\email{vergesclara@gmail.com}
\affiliation{Center for Astrophysics $\vert$ Harvard \& Smithsonian, Harvard University, Cambridge, MA 02138, USA}
\affiliation{Physics Division, Lawrence Berkeley National Laboratory, Berkeley, CA 94702, USA}

\author{A.~G.~Vieregg}
\affiliation{Kavli Institute for Cosmological Physics, University of Chicago, Chicago, IL 60637, USA}

\author{A.~Wandui\orcidlink{0000-0002-8232-7343}}
\affiliation{Department of Physics, California Institute of Technology, Pasadena, CA 91125, USA}

\author{A.~C.~Weber}
\affiliation{Jet Propulsion Laboratory, California Institute of Technology, Pasadena, CA 91109, USA}

\author{J.~Willmert\orcidlink{0000-0002-6452-4693}}
\affiliation{School of Physics and Astronomy, University of Minnesota, Minneapolis, MN 55455, USA}

\author{W.~L.~K.~Wu\orcidlink{0000-0001-5411-6920}}
\affiliation{SLAC National Accelerator Laboratory, Menlo Park, CA 94025, USA}
\affiliation{Kavli Institute for Particle Astrophysics and Cosmology, Stanford University, Stanford, CA 94305}

\author{H.~Yang}
\affiliation{Department of Physics, Stanford University, Stanford, CA 94305, USA}

\author{C.~Yu\orcidlink{0000-0002-8542-232X}}
\affiliation{Department of Physics, Stanford University, Stanford, CA 94305, USA}
\affiliation{SLAC National Accelerator Laboratory, Menlo Park, CA 94025, USA}

\author{L.~Zeng\orcidlink{0000-0001-6924-9072}}
\affiliation{Center for Astrophysics $\vert$ Harvard \& Smithsonian, Harvard University, Cambridge, MA 02138, USA}

\author{C.~Zhang\orcidlink{0000-0001-8288-5823}}
\affiliation{Department of Physics, California Institute of Technology, Pasadena, CA 91125, USA}

\author{S.~Zhang}
\affiliation{Department of Physics, California Institute of Technology, Pasadena, CA 91125, USA}

\begin{abstract}
We use a custom-made calibrator to measure individual detectors'~polarization angles of \mbox{BICEP3}, a small aperture telescope observing the cosmic microwave background (CMB) at 95GHz from the South Pole.
We describe our calibration strategy and the statistical and systematic uncertainties associated with the measurement. 
We reach an unprecedented precision for such measurement on a CMB experiment, with a repeatability for each detector pair of $0.02 \deg$. 
We show that the relative angles measured using this method are in excellent agreement with those extracted from CMB data. 
Because the absolute measurement is currently limited by a systematic uncertainty, we do not derive cosmic birefringence constraints from BICEP3 data in this work. 
Rather, we forecast the sensitivity of BICEP3 sky maps for such analysis.
We investigate the relative contributions of instrument noise, lensing, and dust, as well as astrophysical and instrumental systematics. 
We also explore the constraining power of different angle estimators, depending on analysis choices. 
We establish that the BICEP3 2-year dataset (2017--2018) has an on-sky sensitivity to the cosmic birefringence angle of $\sigma_\alpha = 0.078\deg$, which could be improved to $\sigma_\alpha = 0.055\deg$ by adding all of the existing BICEP3 data (through 2023).
Furthermore, we emphasize the possibility of using the BICEP3 sky patch as a polarization calibration source for CMB experiments, 
which with the present data could reach a precision of $0.035\deg$.
Finally, in the context of inflation searches, we investigate the impact of detector-to-detector variations in polarization angles as they may bias the tensor-to-scalar ratio $r$.
We show that while the effect is expected to remain subdominant to other sources of systematic uncertainty, it can be reliably calibrated using polarization angle measurements such as the ones we present in this paper.
\end{abstract}

\maketitle

\section{Introduction \& Motivation}

Constraining cosmological models requires high-precision calibration measurements of the instruments used to conduct observations.
Polarization-sensitive CMB telescopes like the \bk series of experiments mainly aim at constraining models of inflation, a hypothetical phase of accelerated expansion occurring for a fraction of seconds after the Big Bang.
The polarized CMB signal can also be used to constrain beyond-the-standard-model manifestations of parity violation in the electromagnetic interaction, also known as cosmic birefringence.
While inflation measurements only require relative and overall calibration of the instrument polarization properties, cosmic birefringence constraints call for absolute and individual detector measurements of polarization angles.
This work focuses on the measurement of the BICEP3 polarization angles and their use in constraining cosmic birefringence and inflation.

\subsection{Cosmic birefringence}
\subsubsection{Theoretical motivation}

In the standard model of particle physics, the electromagnetic interaction conserves parity --- the weak interaction, however, does not~\cite{1956Lee,1957Wu}.
Because electromagnetic and weak interaction are unified in the electroweak regime, it has been proposed that electromagnetic interaction may also violate parity through violation of Lorentz symmetry \cite{lorentz-violation}, Faraday rotation due to primordial magnetic fields at recombination~\cite{biref}, or non-standard interactions with dark energy~\cite{dark-energy} or dark matter \cite{dark-matter}.

In that broader context, cosmic birefringence is defined as the Universe being filled with a parity-violating field. 
The interaction of that field with the electromagnetic field therefore also violates parity.
A popular candidate mechanism is the Chern--Simons interaction~\cite{1988turner}, in which the standard electromagnetic field interacts with a pseudo-scalar field $\phi$ through the Chern--Simons term in the \mbox{Lagrangian}: 

\begin{equation}
\frac{1}{4}g_{\phi \gamma}\phi F_{\mu \nu} \Tilde{F}^{\mu \nu},
\end{equation}
where $g_{\phi \gamma}$ is the Chern--Simons coupling term to the electromagnetic tensor $F_{\mu \nu}$.
Pseudo-scalar fields encompass a large class of fields and include in particular axion-like particles, a 
very well studied class of candidate dark matter particles~\cite{2016marsh}.

This coupling results in the rotation of the polarization plane of photons. 
In the case of Chern--Simons coupling, this effect integrates over the line of sight.
One of the best probes to look for cosmic birefringence is therefore the cosmic microwave background (CMB). 
Emitted only 380 000 years after the Big Bang, CMB photons travel through the entire observable Universe before we detect them on Earth. 
Any integrated effect on light polarization is therefore particularly strong on the CMB.
However, intrinsic degeneracies between birefringence and instrumental polarization makes that effect particularly challenging to measure.

\subsubsection{Impact on CMB polarization}
CMB polarization is canonically described using parity-even E modes and parity-odd B modes. Because they are of opposite parity, standard physics predicts that there is no correlation between E and B signals. 
Similarly, there is no expected correlation between the parity-even temperature field T and the B-mode field.
More specifically, in the spherical harmonic domain, the cross correlation function, or power spectrum, of E and B and T and B are predicted to be zero:

\begin{align}
    \mathcal{C}_\ell^{EB} & = 0 \nonumber \\
    \mathcal{C}_\ell^{TB} & = 0,
\end{align}
where $\ell$ corresponds to the angular multipole.

\begin{figure*}
    \centering
    \includegraphics[width=\textwidth]{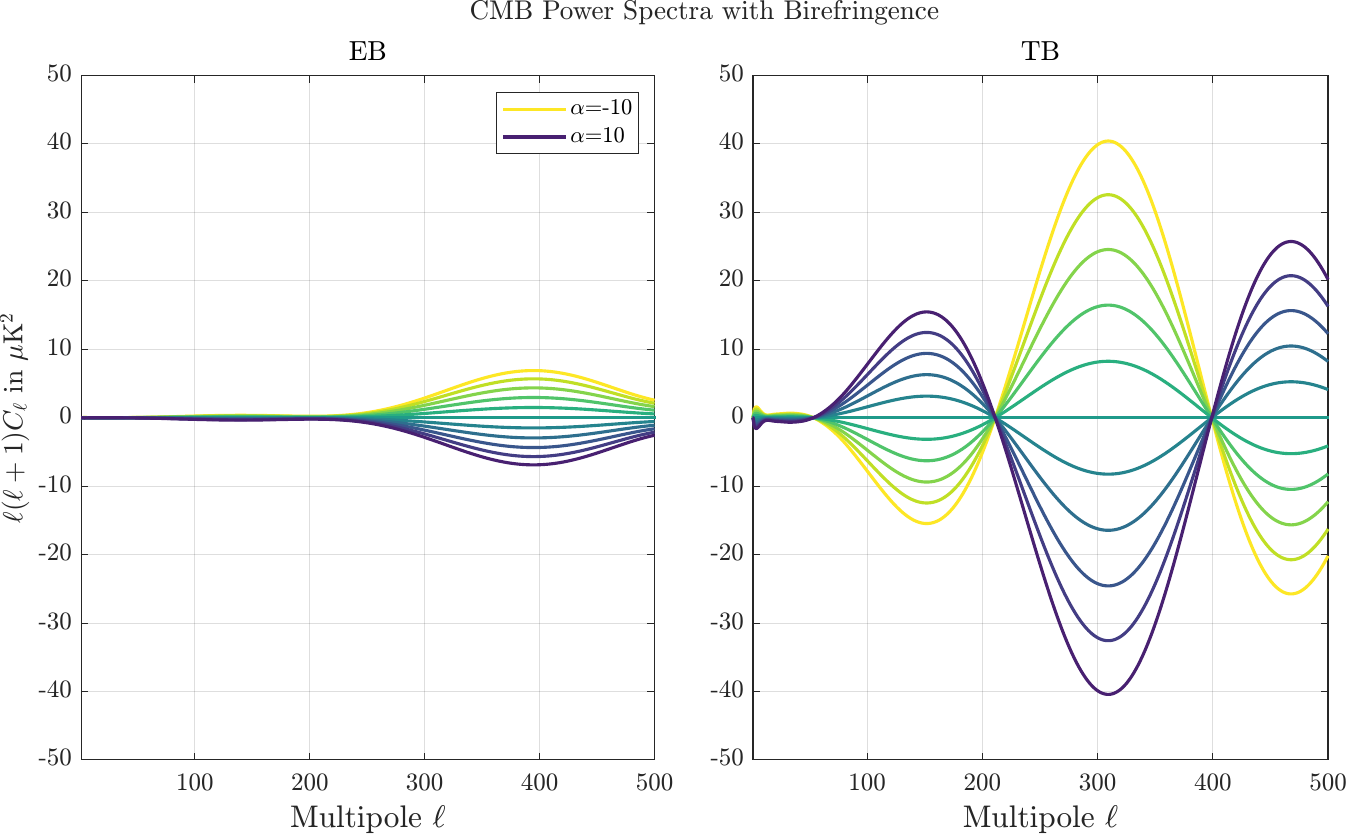}
    \caption{EB and TB power spectrum for a rotation angle $\alpha$ ranging from $-10\deg$ to $10\deg$.}
    \label{fig:ebtb_fid}
\end{figure*}

Because it breaks parity, cosmic birefringence produces a non-zero EB signal. 
This signal can be isotropic or vary over the sky. 
In this paper, we focus on isotropic cosmic birefringence. 
In particular, in the case of the Chern--Simons effect, the isotropic birefringence angle $\alpha$ can be expressed as~\cite{chern-simons}:

\begin{equation}
    \alpha = \frac{\Delta \phi \, g_{\phi \gamma}}{2},
\end{equation}
where $\Delta \phi$ is the change of pseudo-scalar field over CMB photons trajectory, and $g_{\phi \gamma}$ is the coupling constant. The CMB polarization rotation affects all signal types and in particular generates non-zero EB and TB signals, as shown in Fig.~\ref{fig:ebtb_fid}:

\begin{equation}
    \begin{split}
    \Cl^{TT}(\alpha) &= \Cl^{TT} \\
    \Cl^{TE}(\alpha) &= \Cl^{TE}\cos(2\alpha) \\
    \Cl^{EE}(\alpha) &= \Cl^{EE} \cos^2(2\alpha) + \Cl^{BB} \sin^2(2\alpha) \\ 
    \Cl^{BB}(\alpha) &= \Cl^{EE} \sin^2(2\alpha) + \Cl^{BB} \cos^2(2\alpha) \\ 
    \Cl^{TB}(\alpha) &= \Cl^{TE} \sin(2\alpha) \\
    \Cl^{EB}(\alpha) &= \frac{1}{2} \left( \Cl^{EE} - \Cl^{BB} \right)
    \sin(4\alpha) \\
    \end{split}
        \label{eq:alphamodel}
\end{equation}

In principle, any non-zero EB or TB signal measured by a CMB experiment would therefore be a signature for cosmic birefringence. 
However, the cosmic birefringence angle is fully degenerate with the instrument polarization angle.
Any instrument polarization angle that is not calibrated will generate EB and TB power as in Eq.~(\ref{eq:alphamodel}). 
In practice, measured EB and TB signals are precisely used by CMB experiments to calibrate the instrument polarization angle~\cite{keating}, by finding the angle that cancels out the EB/TB pattern.
This operation effectively obliterates any celestial polarization rotation effect, including cosmic birefringence..
To set constraints on cosmic birefringence, one therefore needs an independent determination of the instrument polarization angle.

\subsubsection{Overview of existing measurements}
We first note that this paper uses the IAU sign convention for celestial angles \cite{1973IAU}, resulting in a sign flip when presenting results using the HEALPix sign convention \cite{2005healpix} --- one can refer to \cite{2017alighieri} for more details.
Constraints on cosmic birefringence from CMB data published over the past 15 years are presented in Table \ref{tab:bcconstraints}, alongside with the measurement method that was used.
These measurement methods can be broadly divided into two categories.

The first class of methods relies on the determination of the instrument polarization angles, so that intrinsic instrumental polarization can be taken into account in the analysis. 
This is achieved by observing calibration sources, celestial or artificial, for which the polarization angle is known with a good accuracy and precision. 
The most recent constraint using celestial sources has been placed by the ACTPol collaboration with $\alpha = - 0.07^\circ \pm 0.09^\circ$~\cite{2020choi}.
The systematic uncertainty on that measurement derived from optical modeling is $\sim 0.1^\circ$ \cite{act_syst}.
A number of experiments have also placed limits on an sky rotation angle using artificial calibration sources referenced to gravity. 
However, systematic uncertainties have always dominated the measurement, due to the difficulty of establishing the absolute orientation of the calibrator with respect to the telescope.

To work around that difficulty, a second class of methods has recently been proposed.
It relies on the fact that the Galactic foreground signal is unaffected by polarization rotation from cosmic birefringence, and therefore can be used to disentangle polarization rotation coming from the instrument from a cosmological signal~\cite{minami2019,minami2020-1,minami2020-2}.
It is effectively equivalent to using the Galaxy as a calibration source, assuming its EB signal is either zero or can be modeled.
Such analysis requires multi-frequency information over a broad frequency range, and has been applied to data from the WMAP and \textit{Planck} space missions~\cite{2020minami,2022Diego,2022eskilt-1,2022eskilt-2}.
These various analyses all converge on $\alpha \simeq - 0.30^\circ \pm 0.10^\circ$ (see Table \ref{tab:bcconstraints} for more details).
However, this method suffers from other sources of systematic errors, in particular the incomplete knowledge of foreground polarization in such large sky areas~\cite{2016abitbol,diego2023}.

There is therefore a disagreement between the two classes of methods, which could originate in systematic effects on either or both sides. 
In this work we adopt a calibration-source-based approach, using data taken with the BICEP3 telescope.

\begin{table*}[ht!]

    \centering

\caption{Uniform cosmic birefringence constraints from CMB experiments by order of publication.
The first number is the measurement, then the statistical uncertainty, and in parenthesis the systematic uncertainty if reported.
For ACTPol, we report the value and statistical uncertainties from~\cite{2020choi,2020Namikawa}, and the systematic uncertainty from the follow-up paper~\cite{act_syst}. 
The last two lines show the constraining power of BICEP3 data as presented in this paper, as well as forecasts taking into account additional BICEP3 data and improvements to the RPS setup to mitigate systematics.
\textbf{Constraints originally reported using the HEALPix polarization convention have been sign-flipped to match the IAU polarization convention}.}
\begin{adjustbox}{width={\textwidth},totalheight={\textheight},keepaspectratio}
\begin{tabular}{ ccccc }

\hline
\hline
\textbf{Experiment/Dataset} & \textbf{Frequency [GHz]} & \textbf{$\ell$ range} & \textbf{$\alpha\pm\text{stat}(\pm\text{syst})[^\circ]$} & \textbf{Measurement Method}\\
\hline

\hline
\multirow{2}{*}{QUaD\cite{2009wu}} & 100 & \multirow{2}{*}{200--2000} & $-1.89\pm2.24 (\pm0.5)$ & \multirow{2}{*}{Polarized source}\\
& 150 && $+0.83\pm0.94 (\pm0.5)$ &\\

\hline
BOOM03\cite{2009pagano} & 143 & 150--1000 & $-4.3\pm4.1(\pm0.69)$ & Pre-flight polarized source \\

\hline
ACTPol & 146 & 500--2000 & $-0.2\pm0.5(-1.2)$ & As-designed \\

\hline
WMAP9\cite{2013hinshaw} & 23--94 & 2--800 & $0.36\pm1.24 (\pm1.5)$ & Pre-launch polarized source / Tau A \\

\hline
BICEP2\cite{2013aiken} & 150 & 30--300 & $-1\pm0.2(\pm1.5) $ & Dielectric Sheet \\

\hline
\multirow{3}{*}{BICEP1\cite{2014selfcal}} & \multirow{3}{*}{100+150} & \multirow{3}{*}{30--300} & $-2.77\pm0.86(\pm1.3)$ & Dielectric sheet\\ 
&&& $-1.71\pm0.86(\pm1.3)$ & Polarized source \\
&&& $-1.08\pm0.86(\pm1.3)$ & As-designed \\

\hline
POLARBEAR\cite{2014polarbear} & 150 & 500--2100 & $-1.08\pm0.2 (\pm0.5)$ & Tau A\\

\hline
\textit{Planck}\cite{2016planckparity} & 30--353 & 100--1500 & $-0.35\pm0.05 (\pm0.28)$ & Pre-flight source / Tau A~\cite{2010planckgroundcal,2016planckspacecal}\\
\hline
ACTPol (Choi et al., Murphy et al.)\cite{2020choi,act_syst}& 150 & 600--1800 & $-0.07\pm0.09 (\pm \sim 0.1) $ & Metrology+modeling+point sources \\

\hline
ACTPol (Namikawa et al., Murphy et al. )\cite{2020Namikawa,act_syst} & 98 + 150 & 200--2048 & $0.12\pm0.06 (\pm \sim 0.1) $ & Metrology+modeling+point sources \\

\hline
Planck PR3 HFI (Minami et al.)\cite{2020minami})& 100--353 & 50--1500 & $-0.35\pm0.14$ & Galactic foregrounds \\

\hline
Planck PR4 HFI (Diego-Palazuelos et al.)\cite{2022Diego} & 100--353 & 50--1500 & $-0.30\pm0.11$ & Galactic foregrounds \\

\hline
Planck PR4 HFI + LFI (Eskilt et al.)\cite{2022eskilt-1} & 30--353 & 50--1500 & $-0.33\pm0.10$ & Galactic foregrounds\\

\hline
Planck PR4 HFI + LFI + WMAP (Eskilt et al.)\cite{2022eskilt-2} & 23--353 & 50--1500 & $-0.342 ^{+0.094}_{-0.091}$ & Galactic foregrounds \\

\hline
\hline
BICEP3 2-year (this work) & 95 & 40--500 & $\alpha \pm 0.078 (\pm 0.3)$ & Polarized source \\

\hline
Forecast: BICEP3 7-year + RPS improved performance & 95 & 40--500 & $\alpha \pm 0.055 (\pm \sim 0.07)$ & Polarized source \\

\hline
\hline
\end{tabular}
\end{adjustbox}
\label{tab:bcconstraints}
\end{table*}

\subsection{Inflation}

Constraining inflation through CMB observations calls for extremely high sensitivity measurements of the large angular scale B-mode signal~\cite{1997kamionskowski}.
The impact of polarization angle rotation on the BB signal is shown in Eq.~(\ref{eq:alphamodel}).
As described in the previous section, CMB experiments typically use the spurious EB signal to estimate an instrument polarization angle and correct for this effect prior to estimating B-mode signals.
This approach however only estimates the global instrument polarization angle, as it appears after averaging and weighting of the different detector signals, which all have individual polarization angles.
It therefore does not take into account detector-to-detector variations, which leads to a residual B mode signal that cannot be canceled using this procedure.
This signal, if mistakenly attributed to primordial gravitational waves, could bias estimates of the tensor-to-scalar ratio $r$.

Given the small variability between individual detector angles, such contamination of the primordial B-mode signal is expected to be very low.
Previous studies have found that the bias on $r$ expected from this effect would be $\lesssim 5 \times 10^{-5}$ for the BICEP2 experiment \cite{BKIII}.
However, that result assumed that polarization angles would be distributed randomly on the focal plane, which is not a realistic assumption as we show in Sec. \ref{sec:meas_results}.
Additionally, given the increased sensitivity of CMB experiments aiming to measure $r$ with a significance as low as $\sigma(r) = 5 \times 10^{-4}$, an updated estimation of that effect for a modern CMB experiment like BICEP3 would help alleviate concerns.

\subsection{Dataset \& Methodology}
\label{sec:data_set}
BICEP3 is a CMB telescope continuously observing from the South Pole since 2016 ~\cite{BKXV}.
It targets the CMB signal at degree-angular scales at 95GHz.
Its primary science goal is the search for cosmic inflation by looking for the signature of primordial gravitational waves imprinted in CMB polarization.
BICEP3 focal plane consists of 2400 polarization-sensitive transition edge sensor (TES) bolometers. 
Each detector pair is coupled to a dual-slot antenna array that defines its polarization angle and efficiency.
Additionally, the telescope rotates around its boresight, allowing us to fully reconstruct the polarized sky signal.
BICEP3 maps used in other BICEP/\textit{Keck} analysis, e.g., to constrain primordial gravitational waves~\cite{BKXIII}, are calibrated using E modes through the EB-nulling procedure, and do not use individual detector polarization angle measurements.
Therefore, such maps cannot be used for birefringence constraints.

In this work, we use an artificial, ground-based polarized source to measure the polarization angles of BICEP3 detectors.
We pay particular attention to mitigating and quantifying possible sources of systematic errors in determining these angles.
We use these measurements to simulate two years of BICEP3 data (2017--2018), and report the sensitivity of this dataset for cosmic birefringence constraints.
We also report on the impact of using these measured angles for constraining primordial gravitational waves and inflation.
We choose the years 2017--2018 because data corresponding to these two years have been fully analyzed and used in previous \bk publications~\cite{BKXIII,BKXVI,BKXVII}.
We exclude 2016 because some detector tiles were replaced or moved between the 2016 and 2017 observing seasons, and therefore their polarization angles have not been measured.
The focal plane configuration has remained stable between 2017 and the polarization angles calibration campaign in 2022.

The rest of this paper is organized as follows: in Sec.~\ref{sec:calibration}, we describe the polarization angle measurement and analysis of calibration data, and we present our measurement results.
In Sec.~\ref{sec:analysis}, we detail the methodology of our sky maps analysis, and in Sec.~\ref{sec:cmb_uncertainty} we report the constraining power of BICEP3 data for birefringence searches.
In Sec.~\ref{sec:inflation}, we discuss the impact of this work for inflation searches, and we conclude in Sec.~\ref{sec:conclusion}.

\section{Polarization Angles Measurement}
\label{sec:calibration}

Most of the details regarding the calibration set-up, observations and analysis have been discussed in previous work~\cite{2020cornelison,2022SPIE,cornelison2023}.
Here we summarize the most important aspects of the measurement and subsequent analysis in Sec.~\ref{sec:cal_campaign} and~\ref{sec:rps_analysis}. 
We report our angle measurement results in \ref{sec:meas_results}, and detail our systematic error budget in Sec.~\ref{sec:inst_uncertainty}.

\subsection{Calibration campaign}
\label{sec:cal_campaign}
\subsubsection{Far field observations with BICEP3}
The calibration of instrument polarization angles ideally is performed in the far field of the instrument, so that measurements are most easily
transferred to CMB analysis. 
While many CMB telescopes have far field distances that require observing celestial calibration sources, the small-aperture design and relatively low observing frequency of BICEP3 {\color{black}translates} to a far field distance of $2D^2/\lambda = 171 \mathrm{m}$ (with $D$ the receiver aperture and $\lambda$ the observing wavelength), which allows the use of ground-based sources.

We routinely perform far field calibration campaigns at the South Pole by installing a mirror above the instrument to redirect beams and observe low on the horizon, as the telescope is unable to point below $\sim 45^\circ$ in elevation.
We use a monolithic, honeycomb aluminum mirror, installed directly on the mount --- therefore co-moving with the telescope along the azimuth and elevation axis --- and sitting 1.4m above the center of aperture at an angle of $45^\circ$~\cite{2017Karkare,BKIX,2020StGermaine}.
Calibration sources are then placed atop a 12m mast on a building 200m away from the telescope, fulfilling the far field criterion, and resulting in a source elevation of $2.5^\circ$ above the horizon.

\subsubsection{The Rotating Polarized Source (RPS)}
\begin{figure}
    \centering
    \includegraphics[width=0.45\textwidth]{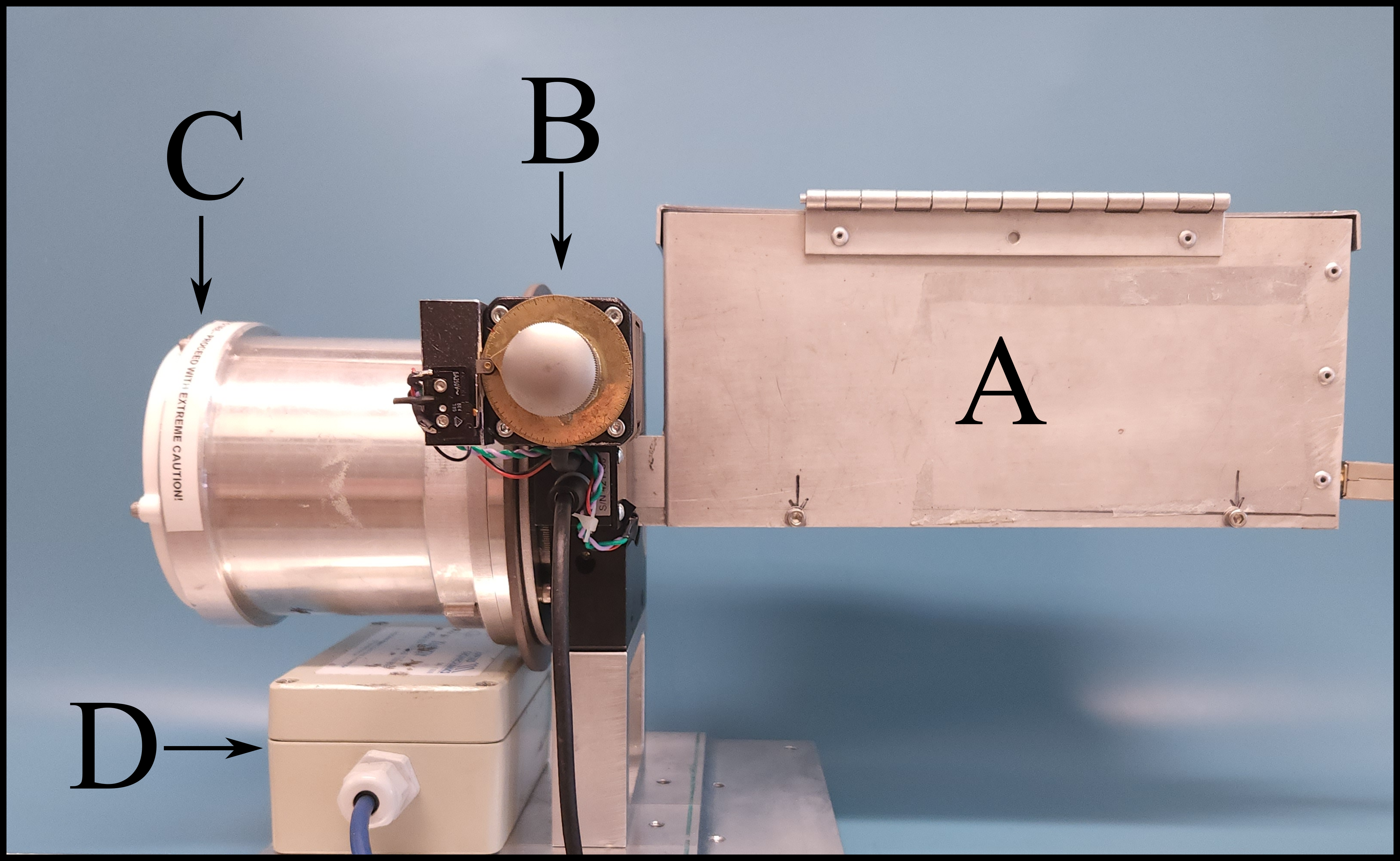}
    \caption{Rotating Polarized Source. An electrically chopped, quasi-thermal noise source (\textbf{A}) is fixed to a rotation stage (\textbf{B}) such that the feedhorn of the source is co-linear with the stage's axis of rotation. We place a wire grid (\textbf{C}) in front of the feedhorn to further polarize the output. We calibrate and monitor the orientation of the wire grid with respect to gravity using a high-precision tilt meter (\textbf{D}). Figure previously published in~\cite{2022SPIE}.}
    \label{fig:rpshardware}
\end{figure}

To determine polarization angles, we observe a quasi-thermal, broad-spectrum noise source transmitting through a polarizing wire grid.
The source is mounted on a rotation stage allowing for $360^\circ$ rotation around the source axis, which allows precise control of the polarization orientation of the linearly polarized signal.
This Rotating Polarized Source (RPS), originally built in 2012~\cite{2012AAS...21942228B}, has undergone many stages of refinements.  
Its use in previous calibrations with BICEP2 and the Keck Array is described in~\cite{BKIV}.
For calibrating BICEP3, the source is configured to transmit between 90-100 GHz and electrically chopped at 20Hz to allow for better background rejection.
A picture of the RPS as used during the 2022 calibration campaign is shown in Fig.~\ref{fig:rpshardware}.
When deployed in the field, the RPS is placed in a temperature-controlled, environmentally-shielding enclosure.

The orientation of the wire grid is established with respect to a surface reference plane, and the orientation of the reference plane during calibration observations is monitored using a high precision tilt meter. 
The control of the orientation of the wire grid, and therefore of the source polarization angle, is of crucial importance in this measurement. 
We detail potential sources of errors as well as mitigating strategies in Sec.~\ref{sec:inst_uncertainty}.

\subsubsection{Observations}

The base element of our scanning strategy is a raster of $9^\circ \times 2^\circ$, covering about 1/60th of the focal plane, during which the RPS angle is commanded to a fixed value. 
Rastering over the source has the advantage of yielding full beam maps for each detector, from which beam parameters (in particular beam centers) can be extracted.
These individual beam parameters are essential in establishing the parameters of the pointing model required to determine individual detector polarization angle.

\begin{figure}[ht]
    \vspace{0.5cm}
    \centering
    \includegraphics[width=0.48\textwidth]{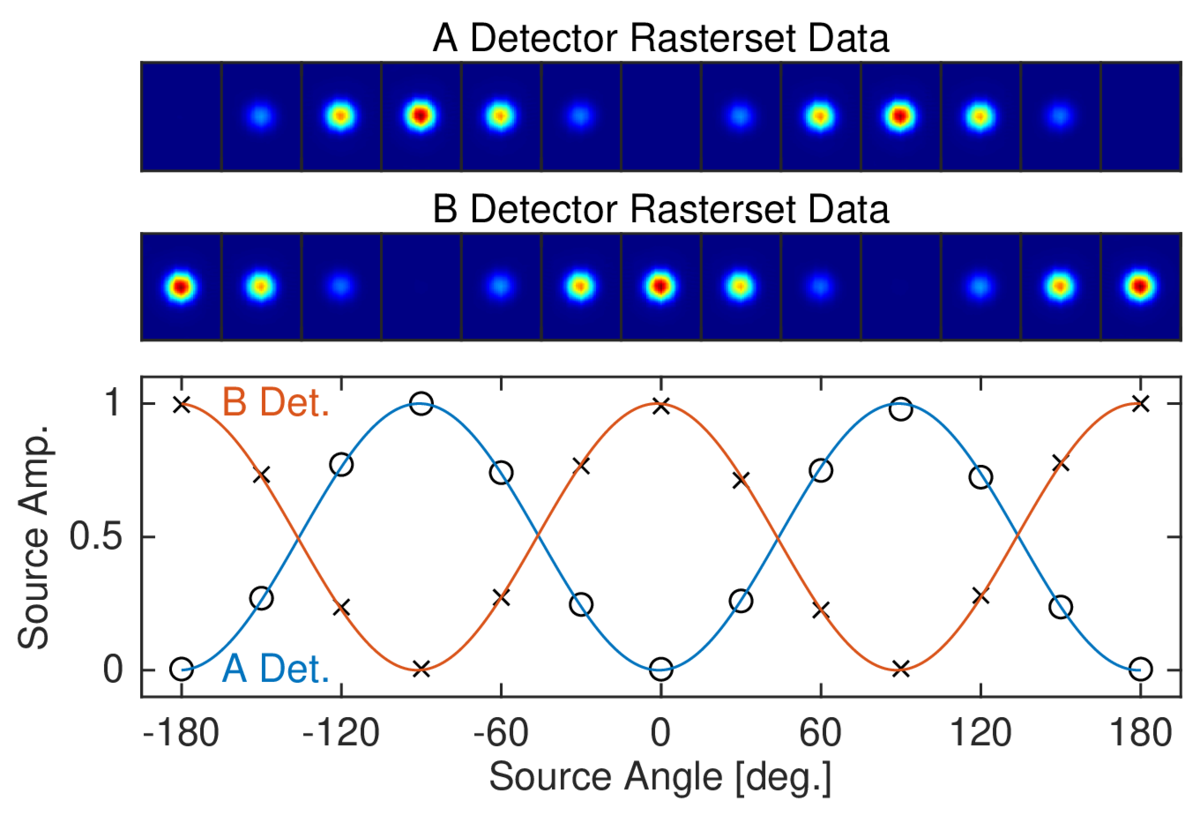}
    \caption{\textbf{Top}: {\color{black}Beam maps of} A and B orthogonal detectors of a given pair, showing modulated beams. \textbf{Bottom}: Resulting modulation curve data
(\textbf{x/o’s}) and fits to the model (\textbf{lines}). Figure previously published in~\cite{BKXV}.}
    \label{fig:modcurve}
\end{figure}

For each focal plane section, we collect thirteen rasters (a rasterset), each at a different RPS angle, evenly spanning from $-180^\circ$ to $180^\circ$ with respect to gravity.
A rasterset thus produces a modulation curve for each detector, as shown in Fig.~\ref{fig:modcurve}, where the beam amplitude depends on the relative orientation between the RPS and the focal plane.
Polarization properties are derived from the modulation in amplitude of the best-fit Gaussian profile as a function of RPS command angle, as detailed in Sec.~\ref{sec:rps_analysis}.

We repeat rastersets as many times as necessary to cover the entire focal plane --- a complete observation takes $\sim 2$ days. 
For this calibration campaign, from 2021, December 24 to 2022, January 28, we were able to collect 10 complete observations of the BICEP3 focal plane, at various telescope boresight rotation angles (also referred to as deck angles). 
We also conducted a number of cross-check observations, including Moon observations, specifically designed to track system stability and probe systematic errors.
A complete description of schedule types can be found in Table 2 of \cite{2022SPIE}.
Our calibration campaign totaled 400 hours of "science" rasterset observations (amounting to 10 full observations of the focal plane), and 290 hours of cross-check observations.

\subsection{RPS data analysis}
\label{sec:rps_analysis}
Our analysis of CMB data uses a standardized
instrument-fixed polar coordinate system, independent of the instrument orientation with respect to the sky \cite{BKXI}.
In this coordinate system, the location of each detector pair is defined with respect to the telescope boresight using coordinates $(r,\theta)$ as shown in Figure 2 of~\cite{BKXI}.
When observing the RPS, the detector polarization angles are obtained as a function of the relative orientation between the RPS and the focal plane. 
We therefore need to relate these angles to the instrument-fixed coordinate system, using information about the orientation of the RPS to gravity, and the pointing model of the telescope.

\subsubsection{Data model}

\paragraph{Parametrization}

\begin{figure}
    \centering
    \includegraphics[width=0.49\textwidth]{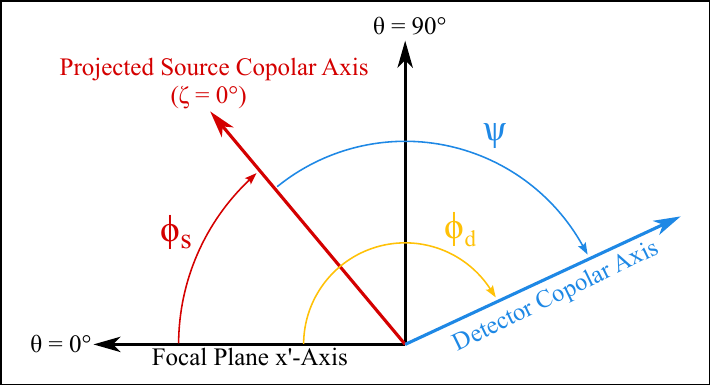}
    \caption{Angles definition in instrument fixed coordinates. The $\theta = 0\deg$ and  $\theta = 90 \deg$ lines define the instrument-fixed, boresight centered coordinate system, as shown for example in Figure 2 of~\cite{BKXI}.
    The $\zeta = 0 \deg$ line is the projection of the RPS zenith on the focal plane.
    The quantity we directly measure is $\psi$, the detector response angle to the RPS signal.  We add to this $\phi_s$, the angle between the focal plane and the RPS zenith, to get the quantity $\phi_d$, the angle of the detector in instrument-fixed coordinates.}
    \label{fig:angles}
\end{figure}

For each detector, we use the following model to fit its modulation curve:

\begin{align}\label{eq:modcurve2}
    A = \,
    G & \times \left(\cos\left[2\left(\psi + \zeta + \phi_s\right)\right]-\frac{\epsilon+1}{\epsilon-1}\right)  \nonumber \\
    & \times \left(n_1\cos(\zeta)+n_2\sin(\zeta)+1\right)
\end{align}
where $A$ is the amplitude of the curve and depends on the following parameters, also schematized in Fig.~\ref{fig:angles}:

\begin{itemize}
       
    \item $\zeta$ the angle of the RPS wire grid with respect to gravity, measured as the angle between the co-polar axis of the grid and the gravitational zenith. Note that in previous work, this angle was referred to as $\theta$, but we have changed it to $\zeta$ to avoid confusion with the coordinate that defines detector pointing in instrument-fixed coordinates $(r,\,\theta)$;

    \item $\psi$ the detector response angle to the RPS signal. It is measured in instrument fixed coordinates as the angle between the co-polar axis of the detector and the projection of the RPS zenith on the focal plane (projection of the $\zeta = 0^\circ$ line);
    
    \item $\phi_s$ the orientation of the focal plane with respect to the RPS, measured from the $\theta=0^\circ$ axis of the focal plane, to the projection of the $\zeta=0^\circ$ line on the focal plane;

    \item parameters not related to the set-up orientation: $G$ the gain of the system, $\epsilon$ the detector polarization efficiency, and $n_1$ and $n_2$ nuisance parameters that account for the RPS miscollimation.

\end{itemize}

$G,\,\psi,\,\epsilon,\,n_1,$ and $\,n_2$ are free parameters that we fit for. 
The other two terms, $\zeta$ and $\phi_s$, are established independently prior to the fit.
$\zeta$ is taken as the command angle of the RPS, corrected by the command angle at which the wire grid is horizontal with respect to gravity.
$\phi_s$ is determined using the telescope pointing model and the orientation of the mirror, using the procedure detailed in~\ref{sec:pointing_model_param}.

Finally, we add $\psi$ and $\phi_s$ to get the detector polarization angle $\phi_d$, in instrument-fixed coordinates:

    \begin{equation}
    \label{eq:phi_d}
    \phi_d = \psi + \phi_s.
    \end{equation}
It is measured from the reference $\theta=0^\circ$ axis on the focal plane, to the co-polar axis of the detector as it is projected onto the sky, as shown in Fig.~\ref{fig:angles}.

\paragraph{Per-pair polarization angles}
Co-located, orthogonally polarized detectors are most often referred to as A and B detectors, which generally align near $\phi_d=0^\circ$ or $\phi_d=90^\circ$.
This designation however is relative to the module wiring, and not to the telescope orientation. 
As such, A and B detectors might have different orientations with respect to an absolute reference depending on how a tile is installed on the focal plane.
Here, we prefer to use the designation of H (horizontal) and V (vertical) for detectors aligned horizontally and vertically with respect to the instrument-fixed x-axis (i.e., $\phi_d=0\deg$)~\cite{2020StGermaine}.

The response for a given detector, $z_d$, to the sky's unpolarized $T$ and polarized components in Stokes $Q$ and $U$ is
\begin{equation}
z_d = T+\gamma(Q\cos2\phi_d+U\sin2\phi_d),
\end{equation}
where $\gamma=\frac{1-\epsilon}{1+\epsilon}$ in the sky coordinate system~\cite{2007Jones}.

The polarized signal in CMB observations is obtained by differencing the time-ordered response of H and V detectors of the same pair:

\begin{equation}
\begin{split}
z_{\text{diff}} &= \frac{z_H - z_V}{2}\\ 
&=  \rho_{pair}\left(Q\cos2\phi_{pair}+U\sin2\phi_{pair}\right)\\
\end{split}
\end{equation}
where $\phi_{pair}$ and $\rho_{pair}$ are the effective per-pair polarization angle and polarization efficiency, respectively:

\begin{align}
    \phi_{pair} & = \, \frac{\phi_H+\phi_V-\pi/2}{2} ... \nonumber \\ 
     & +  \frac{1}{2}\tan^{-1}\left[\frac{\gamma_H-\gamma_V}{\gamma_H+\gamma_V} \times \frac{\cos(\phi_H-\phi_V)}{\sin(\phi_H-\phi_V)}\right]
\end{align}

   \begin{align}
    \rho_{pair} =  \frac{1}{2} & \Bigl[\left(\gamma_H\pm\gamma_V\right)^2\cos^2\left(\phi_H-\phi_V\right) ... \nonumber \\
    & + \left(\gamma_H\mp\gamma_V\right)^2\sin^2\left(\phi_H-\phi_V\right)\Bigr]^{\frac{1}{2}}
    \end{align}
It is the weighted ensemble average of $\phi_{pair}$ over all detectors pairs that is representative of the angle-corrected CMB maps that we create in Sec.~\ref{sec:analysis}. 

\subsubsection{Transfer of coordinate system}

\label{sec:pointing_model_param}
The detector response angles $\psi$ that are fit from the modulation curves are measured with respect to the RPS zenith direction, as we observe the RPS using the mirror.
Ultimately, we want to use these angles as we observe the CMB, and therefore need to transfer that information to our instrument-fixed coordinate system.
As shown by Eq.~(\ref{eq:phi_d}) and Fig.~\ref{fig:angles}, this requires knowing $\phi_s$, the orientation of the focal plane with respect to the RPS.
This can be achieved by comparing beam centers measured during RPS observations to the fiducial experiment beam centers derived from CMB observations. 

\begin{figure*}
    \centering
    \includegraphics[width = \textwidth]{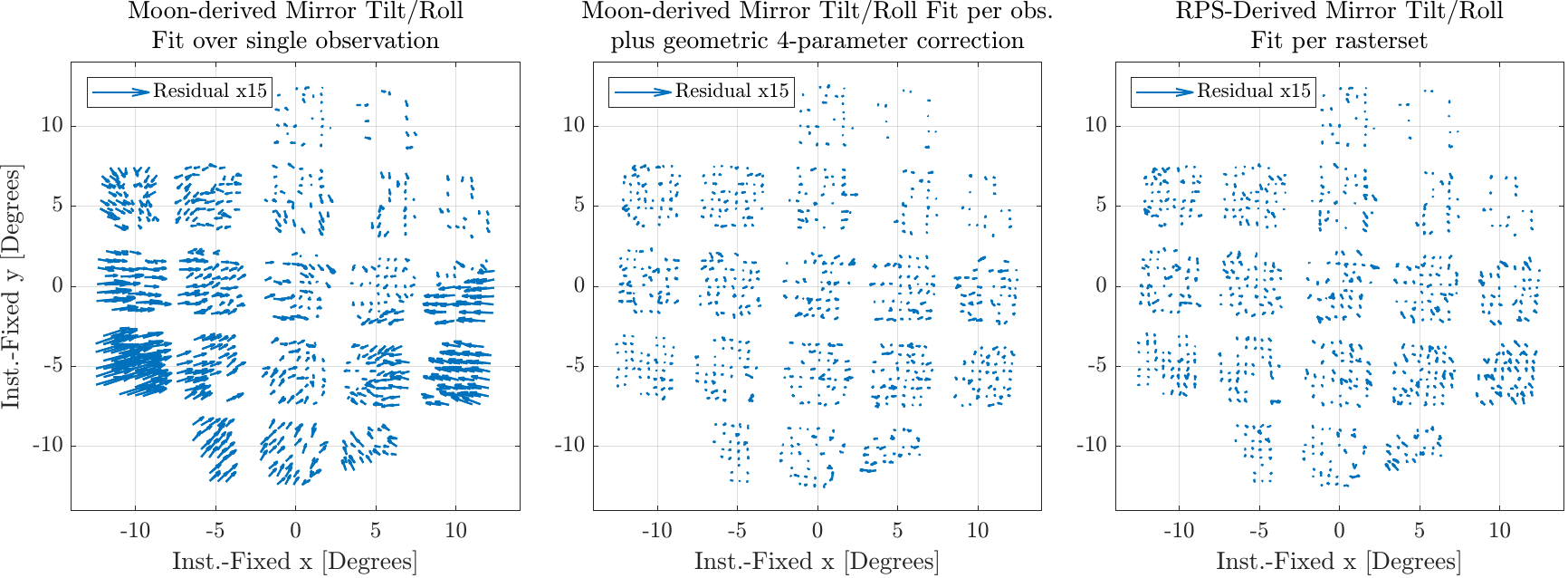}
    \caption[Beam Center Residual Quiver Plots]{Beam center residual quiver plots for one observation, projected from polar coordinates $(r,\theta)$ to cartesian coordinates $(x = r \cos\theta, y = r \sin\theta)$. \textbf{Left}: Residuals when keeping the mirror position fixed to the Moon-derived position for the analysis of the entire RPS observing campaign. \textbf{Center}: Residuals when fitting a global rotation and translation to the initial pattern. \textbf{Right}: Residuals when using RPS data to constrain the mirror position during the RPS observing campaign.}
    \label{fig:beam_cen_res}
\end{figure*}

The instantaneous apparent location and orientation angle of the source in instrument-fixed coordinates, ($x, y, \phi_s$) can be computed from a pointing model, which takes into account all relevant parameters (mount and mirror position, polarization reflection off the mirror, etc.).
In previous work~\cite{2020cornelison,2022SPIE}, we relied on Moon observations to first establish the mirror orientation, parametrized by its tilt and roll.
This mirror orientation can then be used to determine the RPS position.
This method however leaves us with residuals between CMB-derived and RPS-derived beam centers at the level of $\sim$1 arcminute, as seen in the left panel of Fig.~\ref{fig:beam_cen_res}.

The pattern of residuals is coherent, and we find it to be observation-dependent, which excludes a fixed systematic offset, but rather points to short-term fluctuations.
We have evidence from repeated, short ($\sim$1 hour) Moon observations that the mirror is physically moving on such time scales.
This can be attributed to changing meteorological conditions, the sunlight pattern over the mirror changing along the day, etc.
We therefore opt for a data-driven approach where we use the constraining power of the RPS data themselves to account for variations in $\phi_s$ rasterset to rasterset.

The first approach consists in keeping the source position and mirror orientation fixed in the analysis for the entire RPS campaign, and accounting for the remaining apparent offset by fitting an overall rotation angle and translation parameter to the residual pattern for each rasterset.
This approach accounts for any parameter error or missing information in the pointing model in a way that does not explicitly reference the mirror orientation or other physical effects.
The second approach consists in explicitly fitting for the mirror orientation (tilt and roll) for each rasterset --- as they enter as inputs to the pointing model, we can let these parameters be free and fit them to minimize the residual pattern.

As demonstrated in Fig.~\ref{fig:beam_cen_res}, both approaches are effective and perform similarly.
The scatter in the residuals is reduced to 0.53 arcminutes when fitting a rotation plus translation, and to 0.54 arcminutes when fitting the mirror parameters. 
Moreover, fitting for the mirror roll effectively reduces translation and rotation in a way that is both unique to this pointing model parameter, and with the right amplitude to account for the residuals.
This gives us confidence that most of the residual pattern can be accurately accounted for by fitting the mirror parameters per rasterset, i.e., accounting for mirror movement or deformation on short time scales.
We therefore adopt this second approach as our baseline.
The small levels of remaining residuals are indicative of other pointing model uncertainties, that we model and account for in Sec.~\ref{sec:pointing_mod}.

\subsection{Polarization angle results}
\label{sec:meas_results}

We present in this section the final values for BICEP3 individual pair polarization angles derived from the analysis of the entire RPS campaign.

\subsubsection{Polarization angles}
\label{sec:pol_angles}
\begin{figure*}[phtb!]
    \centering
    \begin{adjustbox}{width=\textwidth,keepaspectratio}
        \begin{tabular}{cc}
            \includegraphics{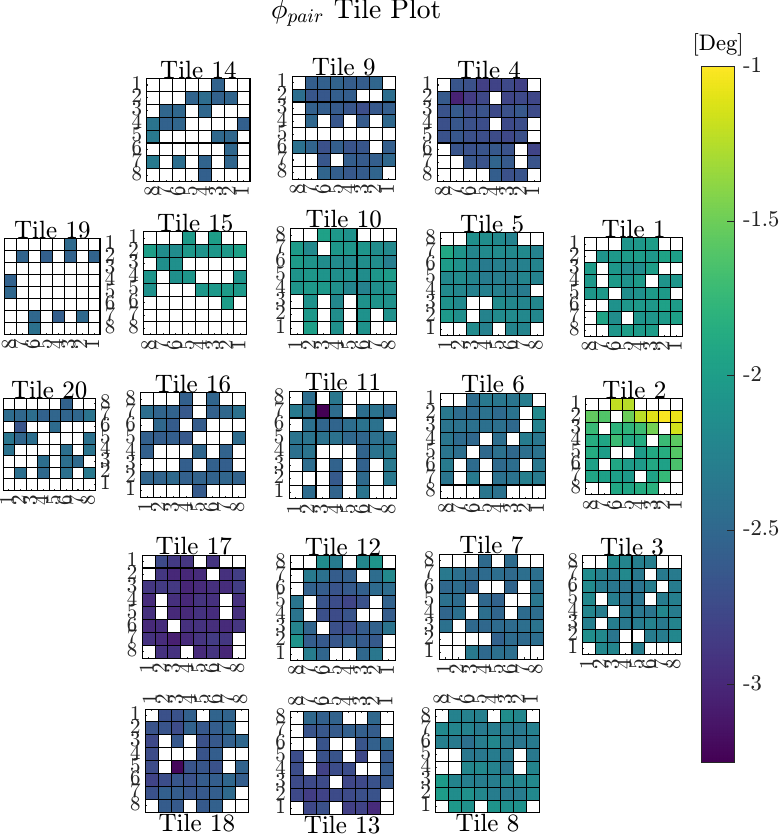} & \includegraphics{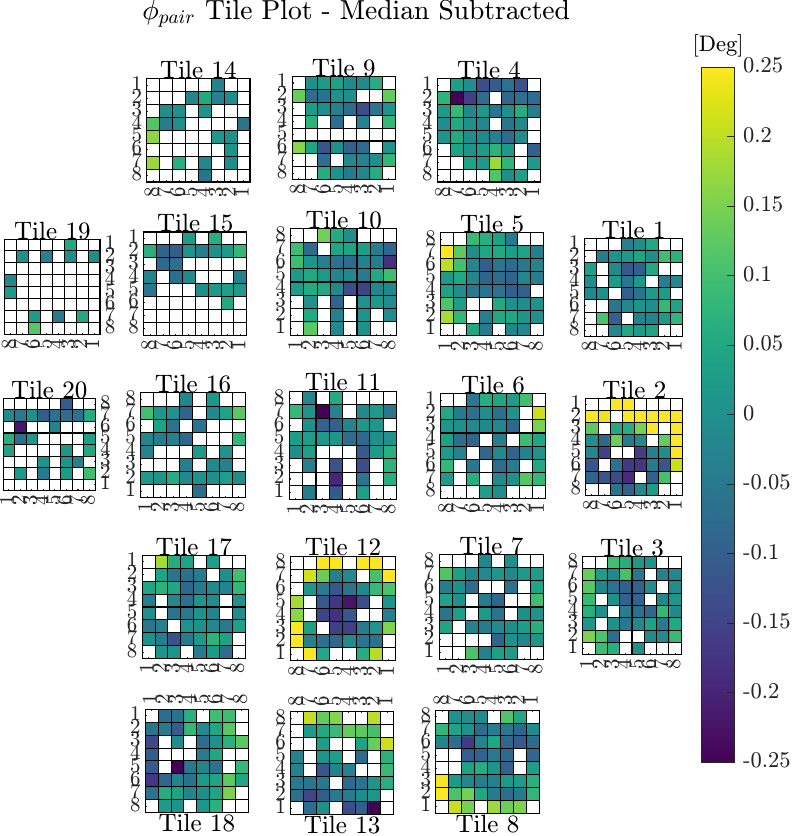}\\
        \end{tabular}
    \end{adjustbox}
    \caption[Calibration Results: Polarization Angles]{Calibration results of polarization angles across the focal plane. The \textbf{left panel} shows the median value of $\phi_{pair}$ for each pair over 10 observations. The pattern over the focal plane is dominated by per-tile effects. In the \textbf{right panel}, the median angle for each tile has been subtracted, and smaller variations within tiles are seen (note the different color scale).}
    \label{fig:rps_results_angles}
\end{figure*}

Per-pair polarization angles $\phi_{pair}$ are shown in Fig.~\ref{fig:rps_results_angles}.
For each detector pair, the value plotted here is the median value over our 10 science rasterset
observations. The overall median polarization angle across the entire focal plane is $-2.4^\circ$.
We note that this value is somewhat arbitrary, since the reference which registers the telescope polarization orientation to gravity is itself arbitrarily set --- it corresponds to the zero point of the boresight rotation encoder, which 
loosely aligns the focal plane to gravity.

Various trends over the focal plane can be identified.
First, we note differences in the per-tile median values on the order of $\sim0.5^\circ$ --- this is what dominates the as-measured values shown in the left panel of Fig.~\ref{fig:rps_results_angles}.
This results from play in the rotational alignment, or clocking, of the modules as they are installed individually on the focal plane.
We discuss this effect further in Sec.~\ref{sec:tile_clocking}.
Variations within tiles can also be identified in the right panel of Fig.~\ref{fig:rps_results_angles}, with some tiles (e.g., tiles 2, 8, or 12) exhibiting a radial pattern.
This effect likely originates in the tile fabrication process.

We are also able to precisely measure detector orthogonality for detectors in the same pair.
For perfectly orthogonal detectors, we expect to find $\phi_H - \phi_V - 90\deg = 0\deg$.
We find deviations from orthogonality at the level of $0.65\deg \pm 0.3\deg$, meaning that our detectors are weakly but consistently non-orthogonal.

\subsubsection{Internal consistency}\label{subsec:rpsstat}
\begin{figure*}[ht]
    \centering
    \includegraphics[width=\textwidth]{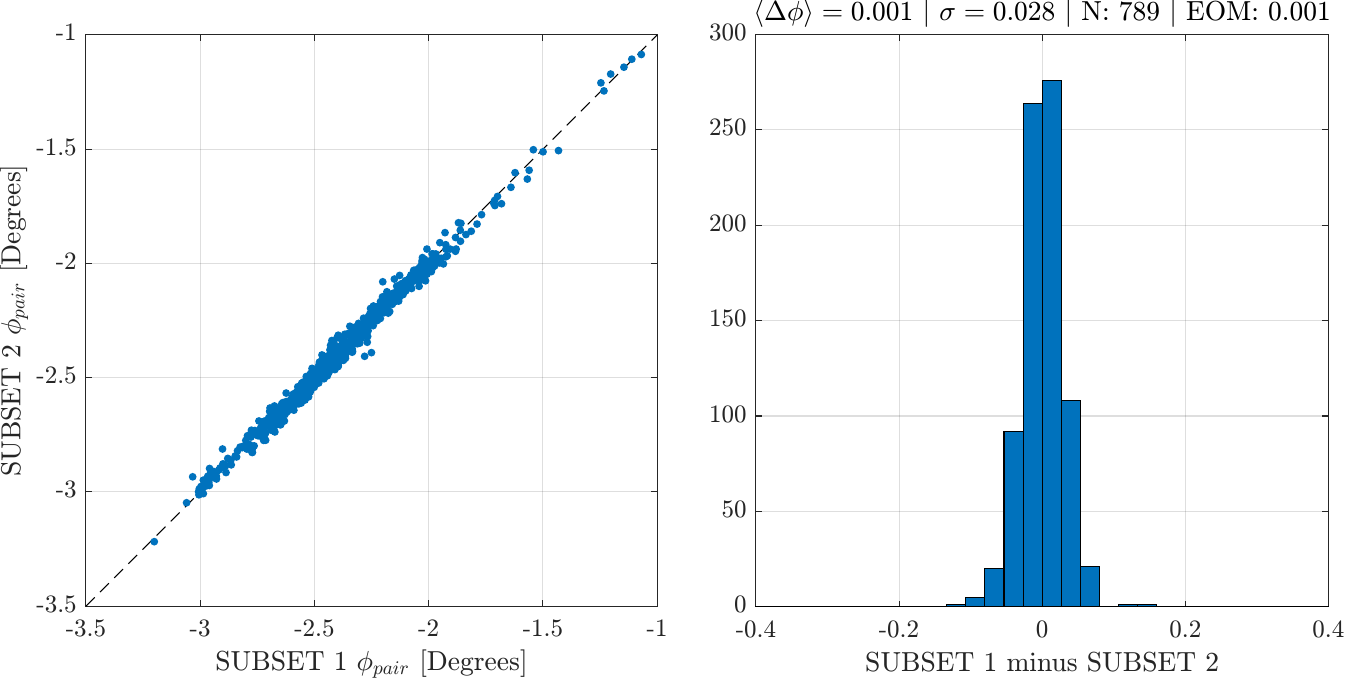}
    \caption[RPS Angle Internal Consistency]{A comparison of two subsets of polarization angles, split evenly across 10 observations. The \textbf{left} plot shows the subsets plotted against each other with each point representing a detector pair. The \textbf{right} plot shows a histogram of the per-pair difference between the two subsets. These plots indicate excellent internal consistency within the data.}
    \label{fig:consistplot}
\end{figure*}

To assess statistical uncertainty and data consistency, we split the 10 observations into two sets of 5 observations and compare them.
The results of this test can be seen in Fig.~\ref{fig:consistplot}.
The correlation between the two subsets is very good, demonstrating a high level of repeatability in these measurements.
More quantitatively, the histogram on the right shows a scatter of $0.028\deg$ when averaging over 5 observations.
When averaging over 10 observations in the final result, we can expect a scatter of $0.028^\circ/\sqrt{2}=0.020^\circ$, which is a good metric of per-pair repeatability.

\begin{figure*}
    \centering
    \includegraphics[width=\textwidth]{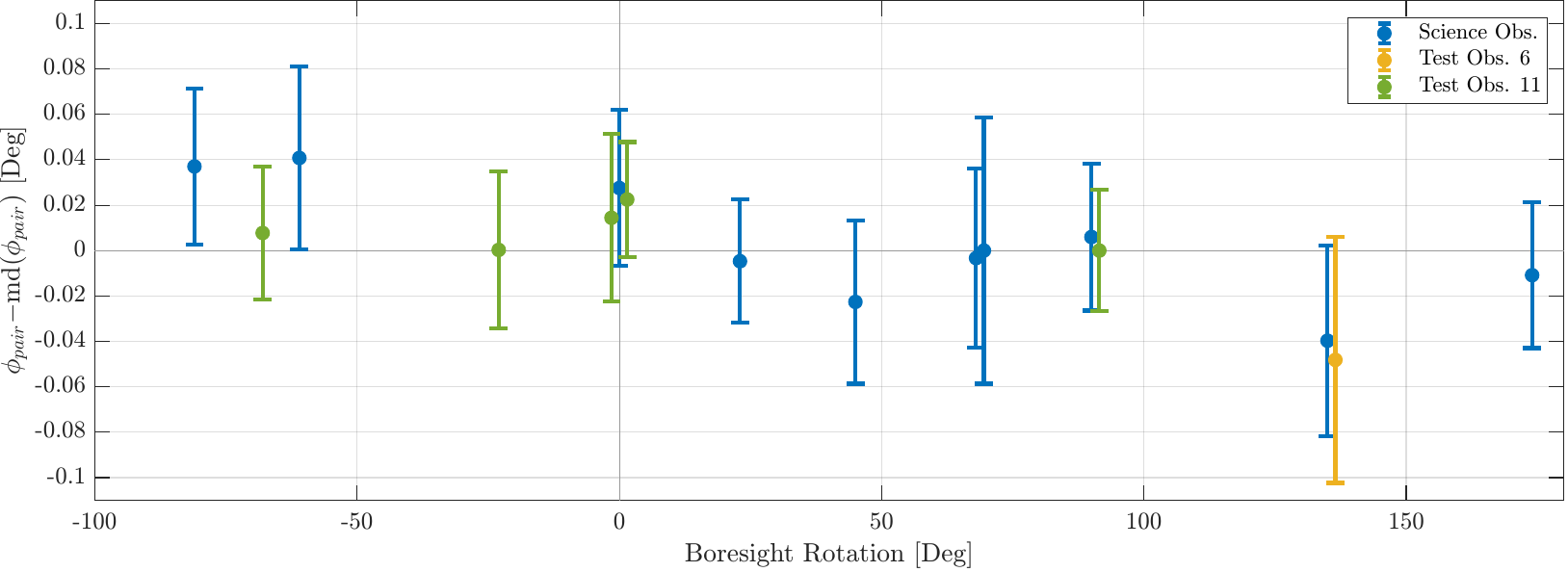}
    \caption[Phi Vs. DK]{Per-pair polarization angles $\phi_{pair}$ across observations at different deck angles. We note a very good repeatability across all observations, including the test observations that had a different scanning strategy --- more details about these test observations can be found in \cite{2022SPIE}. Observations taken at the same boresight rotation angle have been slightly offset for visual purposes.}
    \label{fig:phivsdk}
\end{figure*}

Additionally, for each observation, we look at the deviation from the median as a function of deck angle.
In Fig.~\ref{fig:phivsdk}, we plot the deviations from the median value across observations, i.e., for each pair we plot $\phi_{pair, obs} - \phi_{pair, median}$ where $\phi_{pair, median}$ is taken over 10 observations.
We do this for our 10 science rasterset
observations, as well as test observations that do not necessarily cover the entire focal plane, but still provide a good cross-check while adding some redundancy in deck angle coverage~\cite{2022SPIE}.
For observations taken at the same deck angle, we get a repeatability of $\sim 0.01\deg$, whereas for observations taken at different deck angles, the angle can differ by up to $\sim 0.04 \deg$ from one observation to another.

The per-pair repeatability of $0.02\deg$ established above is a combination of these two effects.
While differences for observations at the same deck angle can be attributed to purely statistical fluctuations, the overall repeatability is affected by systematics as we combine observations at different deck angles.
If systematics did not cancel, the per-pair repeatability could be as high as $0.04\deg$.
However, some of these systematics do cancel or average out as we take measurement at 9 distinct deck angles over almost a full $360\deg$ boresight rotation of the telescope.
This leads to a per-pair repeatability of $0.02\deg$, which is a combination of statistical and systematic effects.

\subsubsection{Polarization angle vs. Module clocking}
\label{sec:tile_clocking}
\begin{figure}[htbp!]
    \centering
    \begin{adjustbox}{width=0.45 \textwidth,keepaspectratio}
        \includegraphics{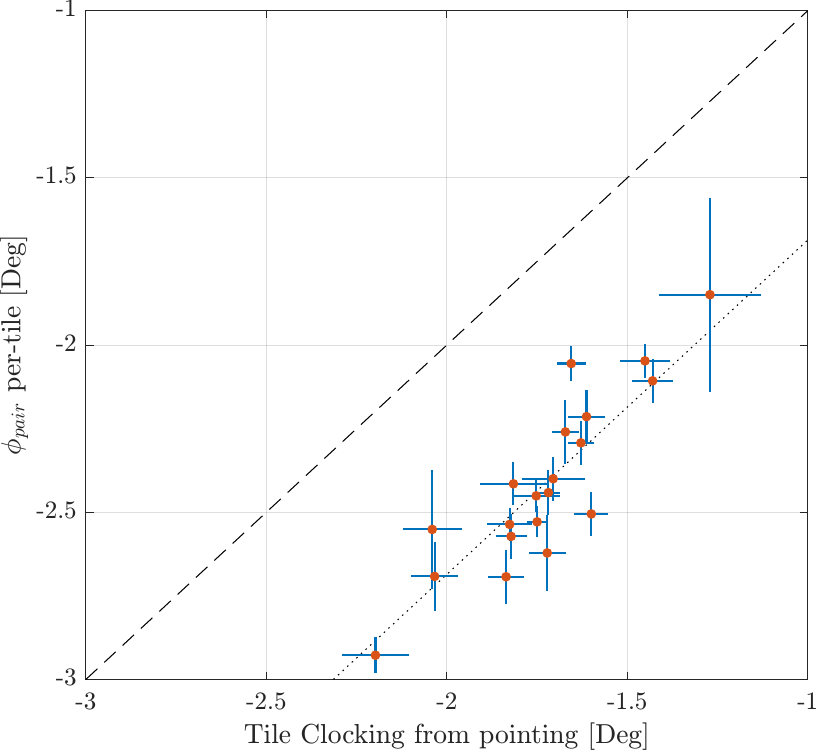}
    \end{adjustbox}
    \caption[Pol Angles Vs. Module Clocking]{Plot of per-tile averages of polarization angles extracted from RPS data vs. module clocking extracted from CMB data, showing {\color{black}agreement in the relative tilt-to-tile offsets} between these two independent measurements.}
    \label{fig:tile_clocking}
\end{figure}

In Sec.~\ref{sec:pointing_model_param}, we took the CMB-derived detector pointings as reference values to establish pointing model parameters.
Ideal pointings are constructed from optical modeling of the detector layout and optics chain, and we measure the achieved pointing by cross-correlating our per-detector CMB {\color{black}temperature }maps with those of the Planck experiment~\cite{BKII}.
One can compare these two sets of pointings (ideal vs CMB-derived) and, after accounting for large-scale transformation across the whole focal plane, we are left with slight per-tile residual rotations due to the imperfect clocking of modules as they are installed on the focal plane.
We expect that clocking of a given tile would become apparent in both the pointing and polarization angle.
Agreement between the two would lend a satisfying cross check of the analysis. To that end, we measure the per-tile rotation angle in the pointing residuals and plot them against the tile-to-tile variation in polarization angle discussed in the previous section.
As shown in Fig.~\ref{fig:tile_clocking}, we do find good agreement {\color{black}in the relative tile-to-tile offsets} between polarization angles and clocking to within $<0.3^\circ$, with $60\%$ of tiles agreeing to less than $0.05^\circ$.

This result is a clear confirmation that the larger tile-to-tile variations in polarization angle are a high-fidelity measurement of the physical clocking between the individual detector modules.
The good agreement of polarization angle measurements with pointing offsets derived from completely independent data and analysis gives us high confidence in the relative values of these polarization angles.

{\color{black}It should be noted that the overall offset of $\sim-0.7^\circ$ between the polarization angles and physical clocking is the value by which our CMB analysis would be biased without the effort of a direct, absolute measurement. 
In addition to containing the significant systematic uncertainty discussed in the following section, this offset also encompasses real physical effects.
The origin of such effects are not exactly known but we believe they arise from non-idealities in our phased-array antennas, which causes the orientation of their polarization sensitivity to deviate from their physical orientation. 
This hypothesis is supported by measured non-orthogonalities between H and V detectors within a given detector pair, consistent in amplitude across all pairs on the focal plane, as mentioned in \ref{sec:pol_angles} and illustrated in Figure 5.4 of \cite{cornelison2023}.}

\subsection{Systematic uncertainty}
\label{sec:inst_uncertainty}
In this section, we focus on how systematic errors contribute to polarization angles measurements.
We approach the systematic error budget from two different perspectives: propagation of uncertainties given a model and some priors in Sec.~\ref{sec:error_analysis}, and laboratory characterization of our calibration apparatus in Sec.~\ref{sec:isaac}.

\subsubsection{Error propagation}
\label{sec:error_analysis}
As shown by Eq.~(\ref{eq:modcurve2}) and Fig.~\ref{fig:angles}, the relevant angles that affect the modulation curves are $\zeta$, the angle of the RPS wire-grid with respect to gravity, and $\phi_s$, the orientation of the RPS with respect to the telescope.
As detailed previously, $\zeta$ and $\phi_s$ are determined prior to fitting detector angles $\psi$ to the modulation curve.
Errors on $\zeta$ and $\phi_s$ will therefore affect the correct determination of $\psi$.
Additionally, even if $\zeta$ and $\phi_s$ were perfectly known, measurement uncertainties affecting other aspects of the modulation curve (e.g., gain variations) could still affect the measurement.

\paragraph{RPS mechanical calibration -- $\zeta$}
\label{sec:rps_cal}
Mechanical uncertainties arising from the calibration of the RPS itself lead to errors on $\zeta$, the angle between the RPS wire-grid and the local gravity.
Such uncertainties can be broken down into three contributions: the command angle of the rotation stage, the angle of the wire-grid with respect to a reference surface (typically the RPS base plate), and the angle of this reference surface with respect to gravity.

To measure these angles, we place the RPS into the vice of a knee mill, ensuring that the RPS reference surface is parallel to the mill translation.
Using a centering  microscope, we count the grid's wires until we find the command angle at which we see no travel in the wires when translating the linear stage of the mill.
We can then estimate errors on each of the three angles listed above, as detailed in \cite{cornelison2023}.
Most systematics affecting the wire-grid calibration are subdominant (smaller than $\simeq 0.01 \deg$), but a couple of effects stand out as a significant contribution to our error budget.

Tiltmeter calibration -- We calibrate the tiltmeter using a machinist level with a precision of $0.001^{\circ}$ placed on the reference surface.
By tilting the tiltmeter to various inclinations with the level attached to it, we derive conversion parameters from angle to voltage returned by the tiltmeter.
We can then propagate the uncertainty on these conversion parameters to the angles that the tiltmeter measures during the campaign.
We estimate a conservative uncertainty on the tilt meter angle of $0.014^\circ$ for the most extreme angles registered during observations.

Rotation stage backlash -- We performed calibration tests both pre- and post-campaign, and discovered a backlash that developed during the campaign.
More specifically, we uncovered play between the gears of the rotation stage resulting in a backlash of $+0.06^\circ$ that had not been measured in the pre-observation checks.
We are currently unable to determine at what point that backlash developed.
We therefore assume that it existed for the entirety of the campaign.
Furthermore, due to the uneven distribution of torque on the rotation stage throughout its $360$-degree rotation, it is not clear whether the polarization angle was preferentially at either end of the backlash ($0^\circ$ or $0.06^\circ$) or randomly distributed between that range.
As such, we add on a conservative $\pm0.06^\circ$ uncertainty to our error budget to account for this phenomenon.

\paragraph{Pointing model -- $\phi_s$}
\label{sec:pointing_mod}
As detailed in Sec.~\ref{sec:pointing_model_param}, pointing model parameters play a key role in the accurate determination of the orientation of the focal plane with respect to the RPS, $\phi_s$.
We model known sources of uncertainty and propagate them through the pointing model. 
Errors are propagated by calculating $\phi_s$ with an ideal value for a given pointing model input, calculating $\phi_s$ again with that input offset by the given uncertainty, and then taking the difference between the two values, $\Delta\phi_s$.
For each parameter, we propagate both a positive and a negative deviation from the fiducial value.
We repeat this for detectors over the entire field of view of the instrument, as detectors further away from the boresight might see a larger effect.
We choose the maximum of all $\Delta \phi_s$ (positive/negative offsets over all detectors) as an upper limit on the uncertainty for each effect that we consider.

We investigate the impact of parameters describing the position and orientation of the key pointing system elements: the source, the mirror, and the telescope mount. 
As input for perturbed parameters, we use priors corresponding to the most extreme deviations observed during the RPS campaign (e.g., mast swaying due to wind that can affect source position), or parameters derived from other analysis such as star pointing that we regularly perform during standard observation to determine mount parameters~\cite{BKII}.
We find that none of the effects that we consider has an individual contribution greater than a $0.007\deg$ uncertainty on the determination of $\phi_s$ -- one can refer to~\cite{cornelison2023} for a more complete description.
When adding all the effects in quadrature, the total uncertainty budget for $\phi_s$ comes up to $0.01\deg$.

We can compare this number to the angle that we fit to the residuals shown in the right panel of Fig.~\ref{fig:beam_cen_res}, which are representative of unmodeled pointing residuals, after the movement of the mirror has been accounted for.
For each rasterset, we compute the residual pattern and fit a rotation angle to it.
The resulting distribution of angles is representative of the total statistical and systematic uncertainty from errors in the pointing model.
The standard deviation on that distribution of angles is $0.012\deg$, in very good agreement with the total uncertainty from error propagation.
This is a confirmation that we have correctly taken into account and/or addressed in the analysis the determination of the pointing model parameters, and gives us confidence that this measurement can be reliably done.
We choose to use the scatter on the distribution of the angles as the total contribution from pointing uncertainties.

\paragraph{Measurement uncertainties}
\label{sec:mod_curve}

This section summarizes measurement uncertainties that affect the modulation curve itself, and can impact the determination of the detectors' angles even if other model parameters were perfectly calibrated.
Before going through a few effects that we found to be significant, we note that we also investigated differential reflection of polarized light on the aluminum mirror, instabilities in the RPS power, and multi-path coupling due to ground reflections, and found all these effects to be negligible in the context of that work (uncertainty much smaller than $0.01^{\circ}$).

Pair anti-correlations -- We find anti-correlations in polarization angle between orthogonal detectors in the same pair that are measured during the same rasterset.
This is due to small phase drifts between the chopped signal as returned by the detectors and the square-wave used for demodulation.
The result of these drifts is a correlated, amplitude-scaling noise in the modulation curves. 
Since orthogonal detectors are $90^\circ$ out of phase, the effect manifests as a phase shift in the curves, corresponding to almost equal and opposite biases on the estimated angles. 
We simulate the effect by adding correlated, amplitude-scaling noise on simulated modulation curves, and show that this results in an increase in per-pair angle uncertainty of $0.02^\circ$, but no significant bias.

RPS Alignment \& Collimation -- The modulation curve described in Eq.~(\ref{eq:modcurve2}) assumes perfect alignment between the RPS and BICEP3, up to a small collimation correction term.
The shape of the modulation curve becomes more complex when considering the projection of polarization in the case where the plane of polarization of the calibrator is misaligned.
It is difficult to sufficiently constrain from the data the extra parameters that account for these projection effects. 
We opt instead to minimize the impact of misalignment by establishing priors on relevant parameters \cite{2022SPIE}.

We control the collimation such that the pointing of the RPS is no more than $1 \deg$ from the axis of rotation of the rotation stage.
The alignment of the RPS with respect to BICEP3 is controlled in azimuth to be $<1^\circ$ and in elevation to $<5^\circ$ away from perfect alignment.
We initially thought that these priors were sufficient to ensure that the impacts of alignment error were subdominant, as our model predicted an error on the angle of $~0.035\deg$. 
However, we found during further testing in a laboratory setting that the impact of misalignment of such amplitude was much greater than expected, as we detail in the coming section.

\subsubsection{RPS characterization}
\label{sec:isaac}
\begin{figure*}[hbtp!]
    \centering
    \includegraphics[width=\textwidth]{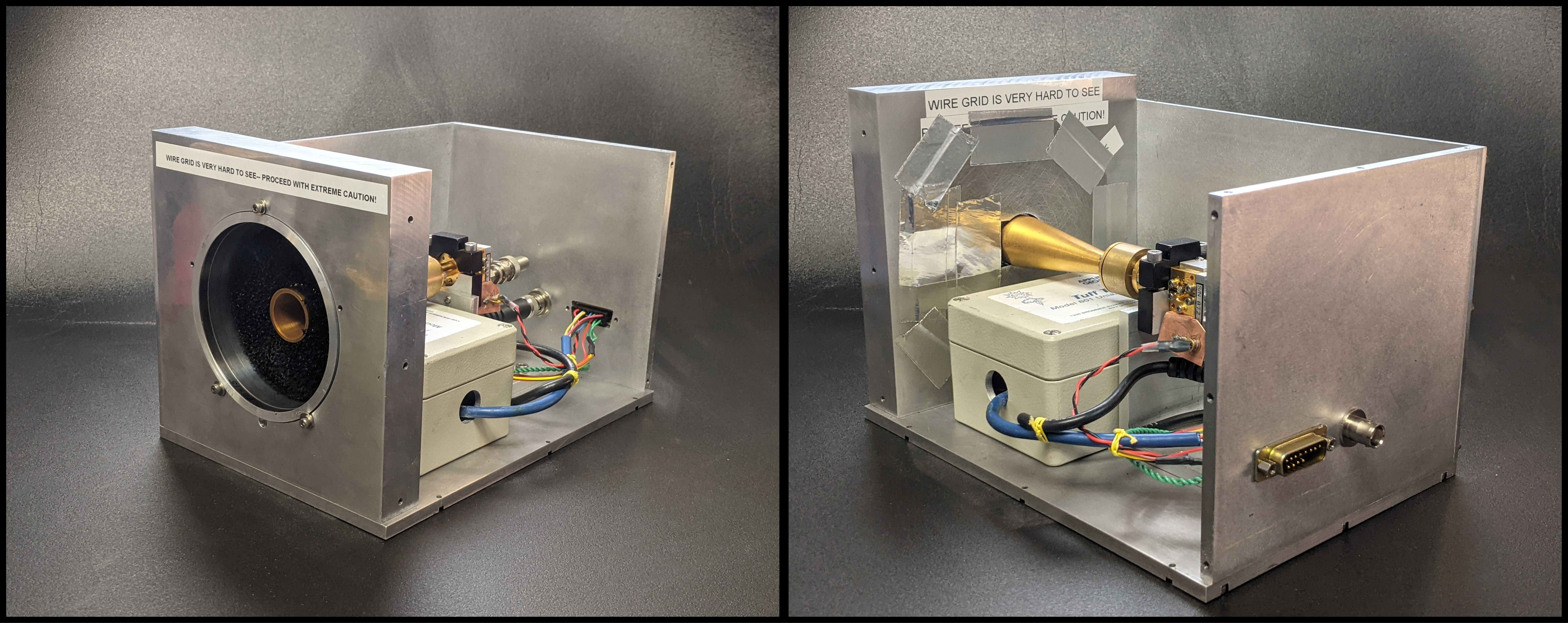}
    \caption[Picture of the ISAAC]{
    Images of the ISAAC benchtop calibrator. Incident radiation is polarized by a wire grid before coupling with a 25~dB corrugated feedhorn. The signal is amplified by 10~dB by a low-noise amplifier before terminating onto a GaAs detector diode. RF absorber enshrouds the feedhorn to mitigate errant reflections from entering the waveguide. Lastly, a TuffTilt tilt meter is used to register the orientation of the wire grid to gravity.
    }
    \label{fig:isaac_pic}
\end{figure*}

\paragraph{The ISAAC}
To verify the predictions of our error propagation model and test the stability of angle estimates, we constructed a compact, room temperature 95~GHz receiver called the In-Situ Absolute Angle Calibrator (ISAAC).
As shown in Fig.~\ref{fig:isaac_pic}, it comprises a polarizing wire grid, a 95~GHz circular feed horn, a Low Noise Amplifier (LNA), and a high-gain detector diode.
Similar to the RPS, the ISAAC is temperature-controlled and contains a tiltmeter to register its wire grid to gravity.
The calibration of the ISAAC wire grid is determined using the same procedure as described in ~\ref{sec:rps_cal}.
Technical specifications and details on ISAAC operations can be found in~\cite{cornelison2023}.

The general approach is to make similar observations of the RPS using the ISAAC as we do with BICEP3 --- the ISAAC output is recorded while the RPS rotates $360^\circ$ in $30^\circ$ increments.
The orientation of the wire grids with respect to gravity of both the RPS and ISAAC are precisely calibrated. 
The angle derived from observing the ISAAC with the RPS should agree with the known ISAAC wire-grid angle to within statistical uncertainty.
Thus, we have a robust and independent confirmation that all aspects of the systematics pertaining to the RPS are well controlled.
Further, by perturbing this benchtop setup in a controlled way, we can impose strong bounds on angle uncertainties we can expect based on our ability to deploy the RPS during observations with BICEP3.

\paragraph{Field measurements}
In-field measurements were taken before the campaign, with the RPS installed in its enclosure and the ISAAC placed $\sim$ one meter away, which places the source in the far field for this apparatus.
These pre-campaign tests were essential in confirming the repeatability of the RPS rotation control, and helped us refine our strategy for homing and zeroing the stage during observations~\cite{2022SPIE,cornelison2023}.
However, they also showed a sensitivity to the relative RPS -- ISAAC alignment along the line of the sight between the two instruments, with small changes in alignment causing variation in the measured angle of up to $0.5^\circ$.
It was hoped at that time that this effect was due solely to the ISAAC, which was a newly built and poorly characterized apparatus compared to the RPS which had been used on several campaigns before.
We therefore decided to proceed with the calibration campaign, and to investigate this effect further once back in a stable laboratory environment.

\paragraph{Laboratory testing}
After the measurement campaign at the South Pole, both the RPS and the ISAAC were brought back to North America, and set up in the lab.
For this laboratory setup, the two apparatus are installed facing each other on a mechanical jig.
Both are able to translate in three orthogonal axes, allowing for precise and repeatable control of distance and alignment.
Further, the RPS is mounted on a horizontal rotation stage to explore sensitivity to azimuthal pointing.

\begin{figure*}[htbp!]
    \centering
    \begin{adjustbox}{width=1\textwidth,keepaspectratio}
    \includegraphics{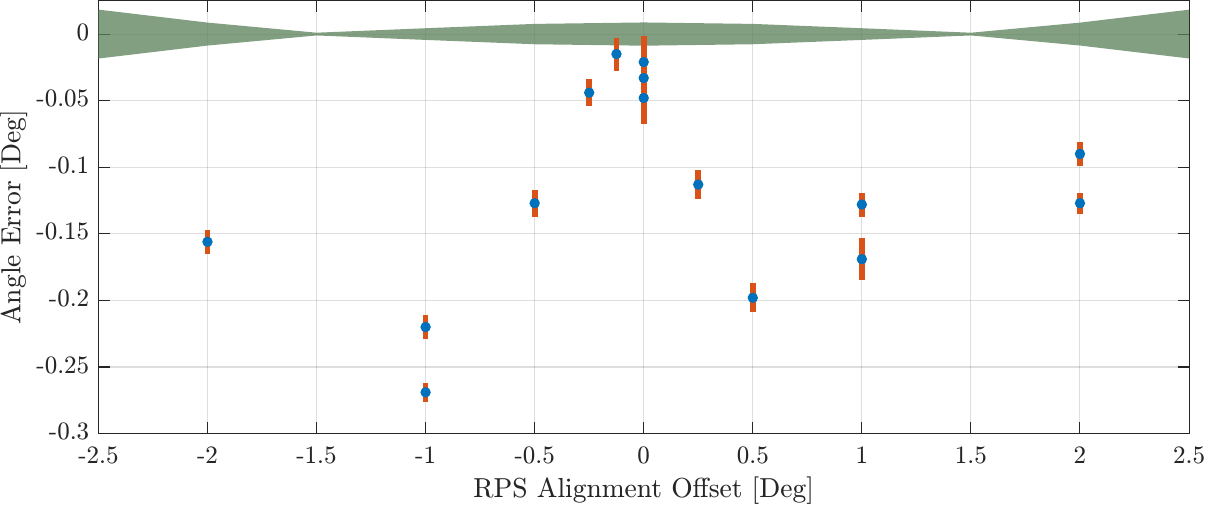}
    \end{adjustbox}
    \caption[RPS-ISAAC Angle Error Vs. Alignment Offset]{
    Scatter plot of the RPS-ISAAC angle error as a function of alignment offset. The error bars are the standard deviations over 10 independent measurements multiplied by an additional factor of 2 for visualization purposes. The green area represents the expected error calculated from geometric projections of polarized light from the RPS~\cite{cornelison2023}. As detailed in Sec.~\ref{sec:isaac}, these deviations have been studied in the lab and their origin traced to multi-path effects and side lobes of the RPS horn interfering with other elements of the setup. The effect will be mitigated for future campaigns.}
    \label{fig:isaac_err_vs_align}
\end{figure*}

We rotate the RPS in azimuth to introduce misalignment between the RPS and the receiver.
At each position, we collect 10 different modulation curves to ensure a good repeatability of the measurement.
As shown in Fig.~\ref{fig:isaac_err_vs_align}, there is a clear dependence of the recovered angle (expected to be zero if the RPS and ISAAC wire-grids are perfectly aligned) vs. the azimuthal alignment offset.
The amplitude of that effect, up to $0.3\deg$, far exceeds the prediction of our model that we had established at $0.035\deg$ for a $1\deg$ misalignment in azimuth.

After numerous tests, we came to the conclusion that the sensitivity to alignment is due to the RPS horn slightly illuminating and diffracting from the conductive aperture of the RPS shroud and external rim of the wire grid. 
This contributes an additional polarized component that is not eliminated by the wire grid and varies across the RPS beam. 
Our RPS design deliberately uses a wide horn (32$\deg$ FWHM) to reduce sensitivity to pointing misalignment, but this has the trade-off of emitting higher power at large angles, requiring extreme care when placing this horn behind a conductive aperture. 
A combination of unexpectedly high sidelobes and placing the horn slightly too far away from the wire grid led to enough illumination of the shroud aperture to disrupt the precision of our measurement.
We have been able to control the amplitude of the azimuth-dependent effect by varying the distance between the horn and the grid, using different horns, as well as adding appropriate baffling and shielding on surfaces which were previously not protected.
While we are confident that these effects can be mitigated by an optimized assembly and a better shielding of the RPS shroud and enclosure, we must however conclude that the alignment sensitivity evidenced by these measurements was present during the 2022 RPS campaign. 
The $0.3\deg$ uncertainty therefore applies to our calibration dataset as a source of systematic uncertainty.

\begin{figure}[hbtp!]
    \centering
    \begin{adjustbox}{width=0.5\textwidth,keepaspectratio}
    \includegraphics{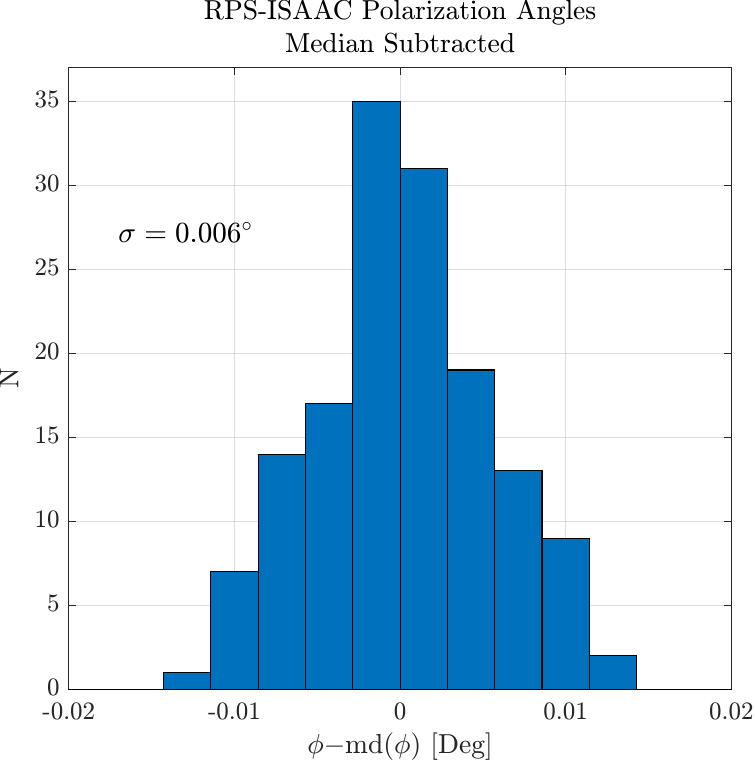}
    \end{adjustbox}
    \caption[RPS-ISAAC Statistical Uncertainty]{
    A histogram of polarization angles derived from measurements between the RPS and ISAAC, demonstrating the small statistical uncertainty of the benchtop measurements.}
    \label{fig:rps_isaac_hists}
\end{figure}

\paragraph{Mechanical repeatability}
Aside from exploring subtle electromagnetic systematics, the benchtop configuration also serves as a way of independently measuring the intrinsic angle repeatability of the RPS.
To do so, we combine the measurements of all 15 measurements discussed above, and subtract the median angle from each distribution, to offset the azimuth-dependent effect. 
As shown in Fig.~\ref{fig:rps_isaac_hists}, we find a resulting scatter of $0.006\deg$, which is in good agreement with the quadrature sum of the mechanical command and homing repeatability that we have established to $0.0054\deg$~\cite{cornelison2023}.
It should also be noted that these measurements are taken with the $0.06\deg$ backlash discussed in the previous section still present.
This shows that it does not significantly contribute to the intrinsic variance of the measurement.

\subsection{Summary}
\label{sec:summary}
We collect all significant sources of uncertainty coming from the RPS measurements in Table \ref{table:rpserrors}.

We quote $\sigma = 0.02\deg$ as our measurement repeatability.
We believe this to be the highest precision achieved to date for calibration of polarization response angles of a CMB telescope.
We have also shown that the relative orientation of detector tiles derived from these data is in excellent agreement with tile orientation derived from a different, completely independent dataset.
This gives us confidence in these relative measurements.

Our systematic error budget is dominated by the azimuth-dependent angle uncertainty at the level of $0.3\deg$.
We note however that even if present during the RPS campaign, such error seems to be stable over time, as evidenced by the small observation-to-observation fluctuations over a month of observations (Fig.~\ref{fig:phivsdk}) and excellent repeatability of per-pair angles (Fig.~\ref{fig:consistplot}).
We expect this effect can be mitigated in future measurements by careful assembly and shielding of the RPS to mitigate electromagnetic effects due to the horn.
The second largest systematic contribution is the rotation stage backlash at the level of $0.06\deg$.
We have already taken steps to mitigate the impact of that effect by installing a high precision optical encoder on the RPS, which can determine the position of the rotation stage up to arcsecond precision~\cite{sjoberg2024}.
We also note that there are other sources of uncertainty for which we can compare our measurement to a model ---  RPS mechanical repeatability and pointing model.
In both these cases, we have demonstrated a good agreement between the measured value and its prediction.
This suggests that we correctly understand the performance and limitations of our current system.
Once dominant effects are mitigated, no other contribution to the systematic error budget would then prevent us from reaching a measurement accuracy at the level of $\mathcal{O}(0.05\deg)$.

Due to the dominant systematic uncertainty that applies to this set of measurements, we cannot assess the validity of the absolute angle measurement.
We therefore decide not to proceed further with birefringence constraints.
The relative angles can be used to forecast the sensitivity of BICEP3 data for constraining cosmic birefringence (Sec.~\ref{sec:analysis} and \ref{sec:cmb_uncertainty}).
They can also inform how polarization angle systematics will impact inflation searches (Sec.~\ref{sec:inflation}).

\begin{table}[h]
\centering
\caption{Sources of uncertainty for the RPS measurements, their raw amplitudes and propagated error on $\phi_{pair}$. The first three categories can be thought of as entirely systematic contributions mostly originating in hardware performance, whereas the last category (measurement repeatability) is a combination of systematic and statistical fluctuations. The error budget is currently dominated by the systematic uncertainty coming from the sensitivity to alignment.}
\begin{tabular}{ lcr }

\hline
\hline

\textbf{Category} & Amplitude & $\sigma(\phi_{pair})$ \\

\hline
\hline

\textbf{RPS performance} & & \\
\hline
Mechanical repeatability & $0.006^\circ$ & $0.006^\circ$ \\
Tiltmeter calibration & $0.014^\circ$ & $0.014^\circ$ \\
Rotation stage backlash & $0.06^\circ$ & $0.06^\circ$ \\
\hline
\hline

\textbf{Pointing model} & & \\
\hline
Focal plane residual rotation & $0.012^\circ$ & $0.012^\circ$ \\
\hline
\hline
\textbf{Measurement uncertainties} & & \\
\hline
Pair anticorrelations & $1.3\%$ & $0.019^\circ$ \\
Alignment error (model) & $\,1^\circ Az/5^\circ El$ & $0.035^\circ$ \\ 
\textcolor{red}{Alignment error (measured)} & & $\textcolor{red}{\sim 0.3^\circ}$ \\ 
\hline
\hline
\textbf{Measurement repeatability } & &  \\
\hline
Per-pair overall repeatability & & $0.02\deg$ \\
Same deck angle repeatability & & $0.01\deg$ \\
Deck-to-deck repeatability & & $0.04\deg$ \\
\hline
\hline
\end{tabular}
\label{table:rpserrors}
\end{table}

\section{CMB Data Analysis Methodology}
\label{sec:analysis}
We use Eq.~(\ref{eq:alphamodel}) as our model to estimate overall rotation angles from power spectra, with some clarifying modifications.
First, EE and BB have poor constraining power due cosmic variance, and do not give the sign of $\alpha$, so we only consider EB and TB in our analysis.
Second, instead of the continuous parameter $\ell$, our spectra are averaged over a range of $\ell$'s into bandpowers $b$, with 
$\Delta \ell = 35$~\cite{2010Chiang}.
Additionally, we use the observed TE, EE, and BB spectra instead of some fiducial cosmological model to calculate EB and TB.
The observed TE, EE, and BB would also be rotated, but at the small angles we are dealing with, it is sufficient to assume rotated and non-rotated TE, EE, and BB are equal.

This section covers the details of this procedure in the context of the analysis of BICEP3 data for the 2017 and 2018 observing seasons --- more details on this dataset can be found in Sec.~\ref{sec:data_set}.
The simulation framework to go from time-ordered data to power-spectrum is outlined in Sec.~\ref{sec:cmb_sims}, and the details of the angle estimation procedure are provided in Sec.~\ref{sec:angle_est}.
Section~\ref{sec:uncal_vs_cal} spells out the analysis choices unique to this study, and Sec~\ref{sec:delta_alpha} concludes with a blind consistency check on real data demonstrating the end-to-end performance and reliability of our analysis.

\subsection{Simulations}
\label{sec:cmb_sims}

\subsubsection{Framework}
Both signal and noise simulations are created at the timestream level.
We use the usual BICEP/\textit{Keck} simulation framework and data model, which includes lensed-$\Lambda$CDM, noise, and Gaussian dust, as described in~\cite{BKXIII}.
We use the pointing information from a given observation to convert an input map into time-ordered data (TOD).
From there, the timestreams are processed through our analysis pipeline the same way as the real data, including deprojection of beam effects, filtering, ground subtraction, etc. 
Timestreams are then binned into maps, from which power spectra can be computed.

In the context of this work, we reprocess and simulate data using only detectors for which we have measured polarization angles, corresponding to about $\sim 90 \%$ of BICEP3 detectors nominally used in the CMB analysis.
The choice of polarization angles assumed at different steps of the simulation and analysis pipeline matters in a way that it does not for the standard CMB analysis.
In particular, our simulation pipeline allows us to specify two sets of angles, the first one used during the TOD generation, and the second one to be used at the map-making step.
In both cases, we can specify angles for each individual detector pair.
The different choices in using the RPS measured angles and their implication for this work are detailed in Sec.~\ref{sec:uncal_vs_cal}.

\subsubsection{Bandpower covariance matrix}
Correctly estimating the bandpower covariance matrix (BPCM) from simulations requires having a sufficient number of noise simulations compared to the number of observables~\cite{2022Beck}.
For our mainline \textit{r} analysis, we create 499 realizations, which is a compromise between minimizing uncertainty on the diagonals of the BPCM and the required computational time and storage space.
For this analysis, we create only 50 simulations, and opt to use the BPCM created for our mainline analysis, under the assumption that the overall statistical properties of the simulations are the same.
We find that this still adequately allows us to complete the analysis while saving a significant amount of processing time and storage space.

\subsubsection{Bandpower window functions}
\label{subsec:bpwf}
Bandpower window functions (BPWF) are a collection of multiplicative factors in harmonic space that produce as-measured bandpowers from theoretical predictions, for a given experiment and analysis choices.
They take into account filtering and signal suppression, from, e.g., inverse-variance apodization and beam convolution, as well as harmonic space binning.
Their role is to allow comparison between measured bandpowers --- which have been apodized, filtered, binned, etc. ---  and a theory power spectrum, so that cosmological parameters can be accurately extracted.
For computational expediency, we use existing window functions created for BICEP3's three-year dataset~\cite{BKXIII,BKVII,2019willmert}.
Doing this introduces a minor source of inconsistency: this analysis uses a two-year dataset with a smaller detector count, which would slightly alter the BPWFs.
Additionally, BPWFs should ideally be calculated for each cross-spectrum $D_\ell^{XY}$, where $X,Y \in {T, E, B}$.
However, since EB signals were not typically included in our mainline B-mode analysis~\cite{2023lau}, the EB BPWFs were not computed individually but were approximated using the geometric mean of the EE and BB BPWFs.
This procedure is known to introduce a few percent error on the final power spectra for EB.

We correct for both these effects by running several sets of simulations that have known input polarization rotation angle. 
We then derive a correction factor to match the output angle with the input one, and we consistently apply this factor throughout our analysis.
This approach is equivalent to re-calculating BPWF, since these are nothing more than a correction factor applied to the power spectra directly. 
Our approach effectively consists in applying this correction factor one step further, once polarization angles have been estimated.

\subsection{Angle estimator}
\label{sec:angle_est}
\subsubsection{Generic formalism}
The model used to estimate the isotropic sky rotation is the one given by Eq.~(\ref{eq:alphamodel}) for EB and TB.
We use the observed TE, EE, and BB spectra instead of a fiducial cosmological model to calculate expectation values for EB and TB.
These observed spectra would also be rotated, but this effect is negligible in the small angle limit.

We can therefore re-express our data model as:

\begin{equation}
    \begin{split}
    \rCb^{TB} &= \oCb^{TE} \sin(2\alpha) \\
    \rCb^{EB} &= \frac{1}{2}\left( \oCb^{EE} - \oCb^{BB} \right)
    \sin(4\alpha) \\
    \end{split}
        \label{eq:alphamodelEBTB}
\end{equation}
where $\oCb$ are the real bandpowers and $\rCb$ are expectation values.
We consider our best-fit angle estimate as the angle which minimizes
\begin{equation}\label{eq:alphachisq}
\chi^2 = \left( \oCb^{XY} - \rCb^{XY} \right)^t \left( \Ccov^{XY}
\right)^{-1} \left( \oCb^{XY} - \rCb^{XY} \right),
\end{equation}
where ${XY}$ can in principle be EB, TB, or both, and $\Ccov$ is the bandpower covariance matrix.
When including both EB and TB, we do include EBxTB correlations in the bandpower covariance matrix.

\subsubsection{Choice of power spectrum}
\begin{figure*}
    \centering
    \includegraphics[width = \textwidth]{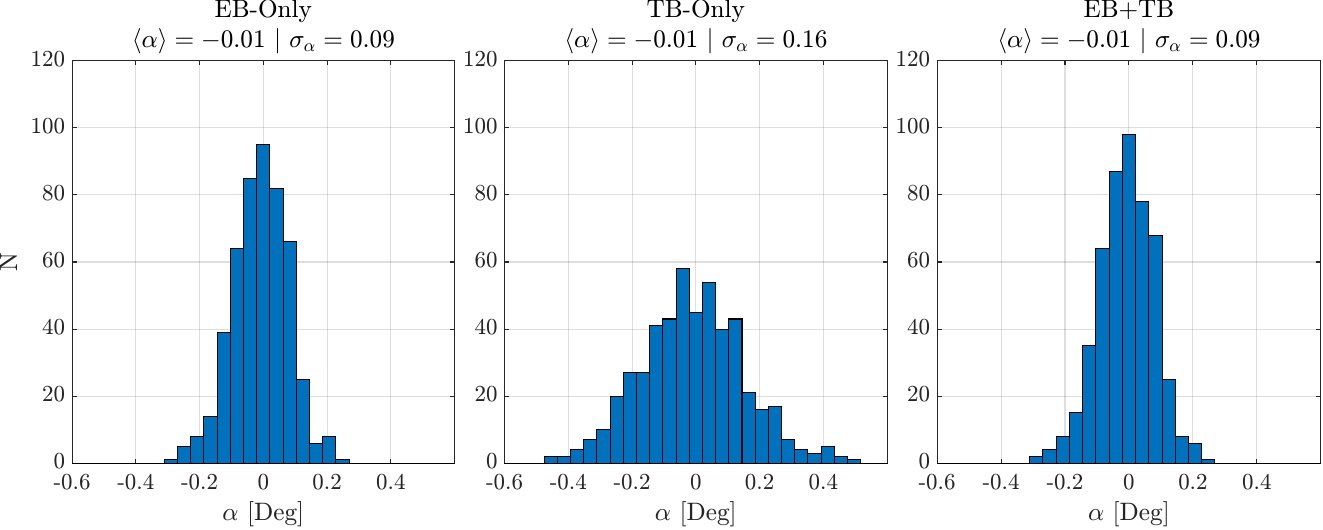}
    \caption[\textwidth]{Histograms of angle estimates from CMB power spectra with angles fit to EB-only, TB-only, and a combined EB+TB estimator. Because TB contributes no additional information to the fit, only noise, the combined EB+TB estimator has roughly the same scatter on $\alpha$ as the EB-only fit.}
    \label{fig:EBTBhists}
\end{figure*}

Since we have two different spectra with similar constraining power, we would ideally combine them in a joint estimator to leverage their constraining power.
However, we show that the TB spectrum contributes no additional information when combining EB and TB together into a single estimator.
This can be demonstrated analytically, as detailed in 
Appendix~\ref{sec:appendix_ebtb}.
We also verify this empirically on simulations, as shown in Fig.~\ref{fig:EBTBhists}.
The EB+TB case has the same scatter on $\alpha$ as the EB-only case, confirming the analytical result.
The uncertainty on the recovered angle can not be improved below that of EB only.
We therefore decide to use EB only in this analysis.

\subsubsection{Choice of multipole range}
\label{sec:ell_choice}
In our standard CMB analysis, we use 9 bandpowers ranging from $\ell \approx 40$ to $\ell \approx 320$.
The lower angular scales are excluded because most of the signal is effectively suppressed due to filtering. 
Higher bandpowers are typically excluded because they are noisier, more subject to beam uncertainties, and their constraining power on $r$ is limited compared to lower multipoles.

In this analysis, we have strong motivation to include higher bandpowers than those used in our standard analysis framework.
In particular, the EB signal peaks at $\ell \approx 400$ and its amplitude is much higher for $\ell > 250$, as shown in Fig.~\ref{fig:ebtb_fid}.
Despite noisier data at high multipoles, we find that we still have good constraining power up until $\ell \approx 500$, with a peak at $\ell \approx 350$.

\begin{figure*}
    \centering
    \begin{adjustbox}{width={\textwidth},keepaspectratio}
\includegraphics{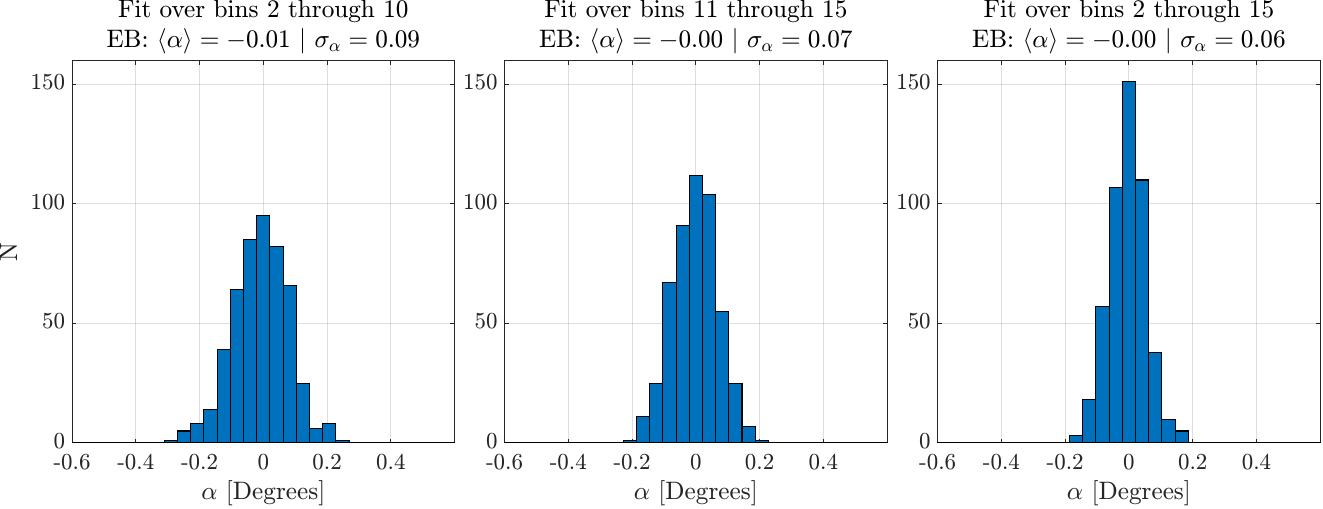}
    \end{adjustbox}
    \caption[Angle Constraining Power Vs. $\ell$]{Demonstration of the constraining power on $\alpha$ when fit over different multipole ranges. Adding 5 additional bins compared to the mainline CMB analysis (bins 11 to 15) increases the constraining power on $\alpha$ by 30\%.}
    \label{fig:constrain_vs_ell}
\end{figure*}

Fig.~\ref{fig:constrain_vs_ell} shows distributions of angle estimates from CMB simulations fit over different multipole ranges.
When comparing the statistical uncertainty on the best-fit angle between our standard 9 bandpowers ($40<\ell<320$) to the scatter when adding 5 more bandpowers ($350<\ell<500$), we find that over 50\% of the constraining power for this analysis is contained in higher multipole ranges.
When including all 14 $\ell$-bins in our EB fit, we find a $30\%$ improvement in angle uncertainty.

We find these insights to be sufficient motivation to report results both using our standard 9 bandpowers as well as results including an additional 5 bandpowers. 
Later on, we refer to these as the 9-bandpower case and 14-bandpower case.
Because the 14-bandpower case has a higher constraining power, we decide to showcase it as our main result.
Including higher multipoles in the analysis calls for a particular attention to beam systematics and uncertainties, that we investigate in detail in Sec.~\ref{sec:beams}.

\subsection{Uncalibrated vs. calibrated data processing}
\label{sec:uncal_vs_cal}
When processing simulation and real data, we have a choice over how to apply calibration information (polarization angles) to the data.

The first, natural approach we take in this analysis is to re-process existing data using RPS-derived angles at the map-making step, as well as creating simulations with RPS-derived angles for both TOD generation and map-making.
In the latter, the data products are calibrated --- since polarization angles are consistently used for making TOD and maps --- and the expectation value for the instrument angle goes to zero.
In the following, we refer to this approach as ``calibrated simulations".

Another option is to use RPS-derived angles to generate TOD, but ideal angles at the map-making step.
This is closer to the state that our (non-reprocessed) CMB data are, since we observe the sky with the true instrument angles, but then assume ideal angles in map-making.
In that approach, the data products remain uncalibrated.

We could in principle use these uncalibrated spectra as is in the analysis.
However, because they are uncalibrated, the expectation value for the EB signal is non-zero, which makes it more prone to multiplicative systematics.
In particular, we identified beam window function uncertainties as a significant source of multiplicative systematic error, as detailed in Sec.~\ref{sec:beams}.
We therefore decide to de-rotate the uncalibrated maps and spectra using the mean of the angles fit to the simulations, and refer to it as ``uncalibrated derotated simulations". 
This approach is therefore quasi-identical to the way we process real data in our standard CMB analysis, with the significant difference that the de-rotation angle comes from RPS measurements and full signal simulations and not from the real data itself.
It also allows the extra insight on any possible systematics imparted on our inflation results from assuming uniform polarization angles -- see discussion in Sec.~\ref{sec:inflation}.

\begin{figure*}
    \centering
    \begin{adjustbox}{width={\textwidth},keepaspectratio}
    \includegraphics{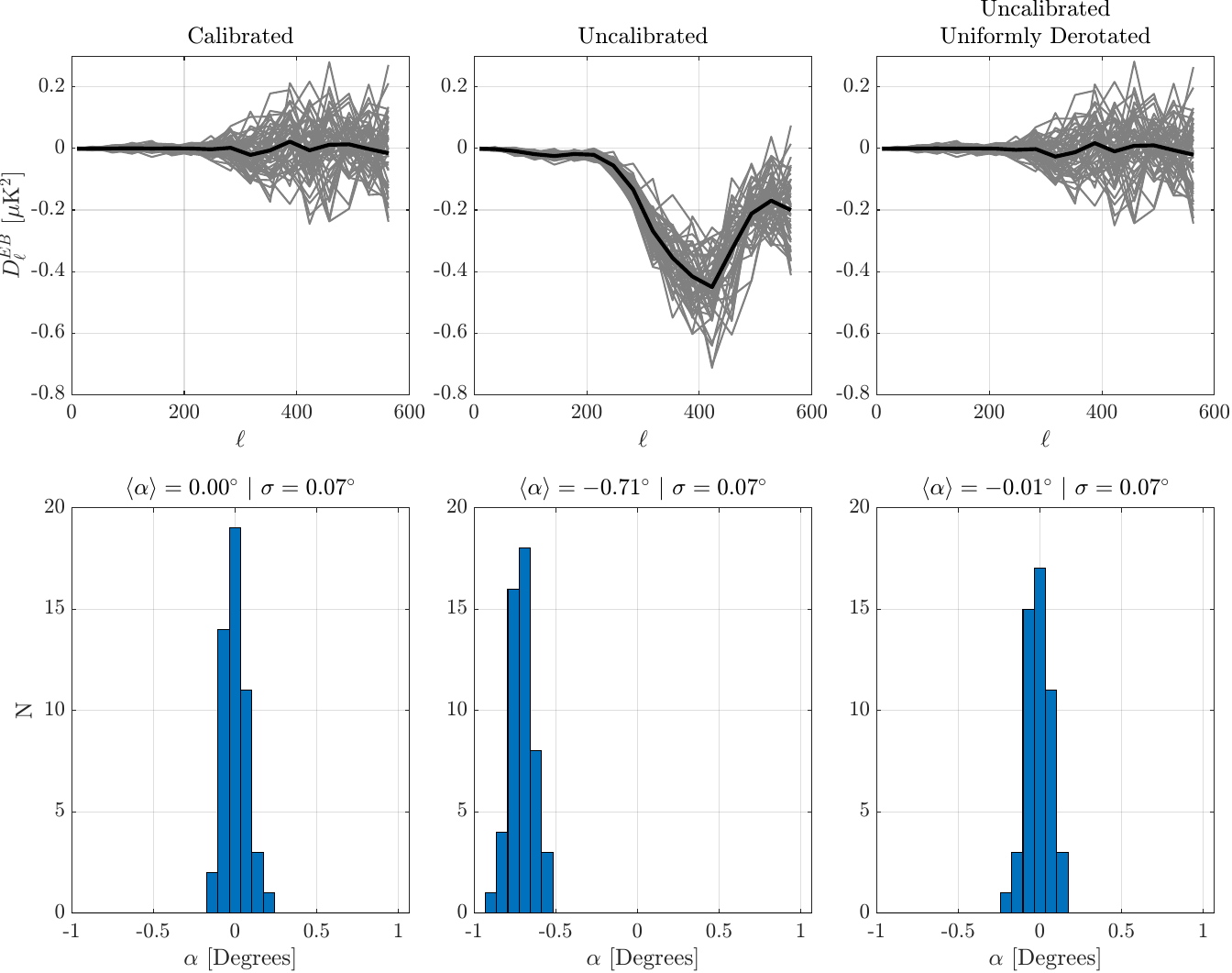}
    \end{adjustbox}
    \caption[Angle-calibrated and uncalibrated Simulations]{Plots of calibrated and uncalibrated $\Lambda$CDM+Noise+Dust simulations, with EB power spectra on the top row and angle best-fits on the bottom row.
    While the scatter on the angles are all the same, the impact of a multiplicative uncertainty from beam window function errors would be greater if EB is non-zero. For that reason, we de-rotate the maps by the mean uncalibrated angle and fit angles to the sims again.}
    \label{fig:angle_corr_uncorr_spectra}
\end{figure*}

Fig.~\ref{fig:angle_corr_uncorr_spectra} shows the cases of $\Lambda$CDM+Noise+Dust simulations including the uncalibrated simulations with and without map derotation.
We show that both approaches have the same statistical power and produce angle estimates identical within the statistical uncertainty.

\subsection{Consistency test on real data}
\label{sec:delta_alpha}

We designed a specific, blind consistency test on real data to build up our confidence that the relative measurements of the polarization angles are correct and correctly taken into account in the analysis.
Inspired by our usual jackknife tests, we use information from RPS calibration to split the data and call this test the $\Delta \alpha$ test.
Note that due to various analysis choices that we have outlined, the dataset we use in this analysis is different from the mainline BICEP3 3-year dataset.
We have therefore also verified that our reprocessed real data pass all the usual jackknife tests~\cite{cornelison2023}.

\subsubsection{\texorpdfstring{Definition of the $\Delta \alpha$ quantity}{Definition of the delta-alpha quantity}}
The angle we fit to real spectra will be a combination of the instrument angle and any celestial isotropic polarization signal:

    \begin{equation}
    \alpha_\text{obs} = \alpha_\text{B3} + \alpha_{sky}.
    \end{equation}
If we split the data into subsets $\{i,j\}$ and subtract the angles fit from those subsets, we are left with only a difference between the instrument angles, as the celestial signals cancel out:

    \begin{equation}
    \begin{split}
    \Delta\alpha &= \alpha_{\text{obs},i} - \alpha_{\text{obs},j}\\
    &= \alpha_{\text{B3},i} + \alpha_{sky} - \alpha_{\text{B3},j} - \alpha_{sky}\\
    &= \alpha_{\text{B3},i} - \alpha_{\text{B3},j}.\\
    \end{split}
    \label{eq:delta_alpha}
    \end{equation}
For the uncalibrated approach, the non-zero angle we get from the EB power spectrum (in the absence of sky signal) comes from the discrepancy between the TODs and map-making angles.
The real BICEP3 data use design values for the polarization angles of the detectors at the map-making step, and all detectors are assumed to have the same polarization angle.
However the angles we measure with the RPS show a detector-to-detector scatter of $\sim1^\circ$.
We can therefore expect a difference between data subsets, i.e., $\Delta \alpha \neq 0$, due to detector-to-detector scatter that is not captured at the map-making step.
In the calibrated approach however, because we use measured RPS angles at the map-making step, we expect these differences between subsets to go away, and we should find $\Delta \alpha = 0$.

The choice of angle subset is arbitrary but limited by the fact that the statistical uncertainty increases with fewer detectors.
In this case, we choose to split the detectors' pairs evenly into thirds.

\subsubsection{Simulations}
We first test this approach on simulations, to be able to judge the significance of the result on real data.
For each subset of angles and each approach (calibrated/uncalibrated), we create 50 simulations and make co-added maps.
We compute power spectra and fit an angle to them using our usual estimator.
Before computing $\Delta \alpha$, we check that the angles fit to the power spectra are consistent with the mean of input angles.
The output is consistent with the input within uncertainty, with deviations being attributed to differences in sky coverage between the subsets.

We then proceed with computing $\Delta \alpha$ on a realization-to-realization basis.
The histograms on Fig.~\ref{fig:dalphahists} show the case  ``Low subset - High subset'', for which the differences are the largest. 
We confirm that the difference between subsets goes away between the uncalibrated and calibrated approach in a way that is statistically significant {\color{black}$\left(3.1\sigma\right)$}. 
We obtain similar results for the other subsets.
This gives us confidence that we can use this test on real data.

\subsubsection{Result on real data}
\begin{figure*}
    \centering
    \begin{adjustbox}{width={\textwidth},keepaspectratio}
    \includegraphics{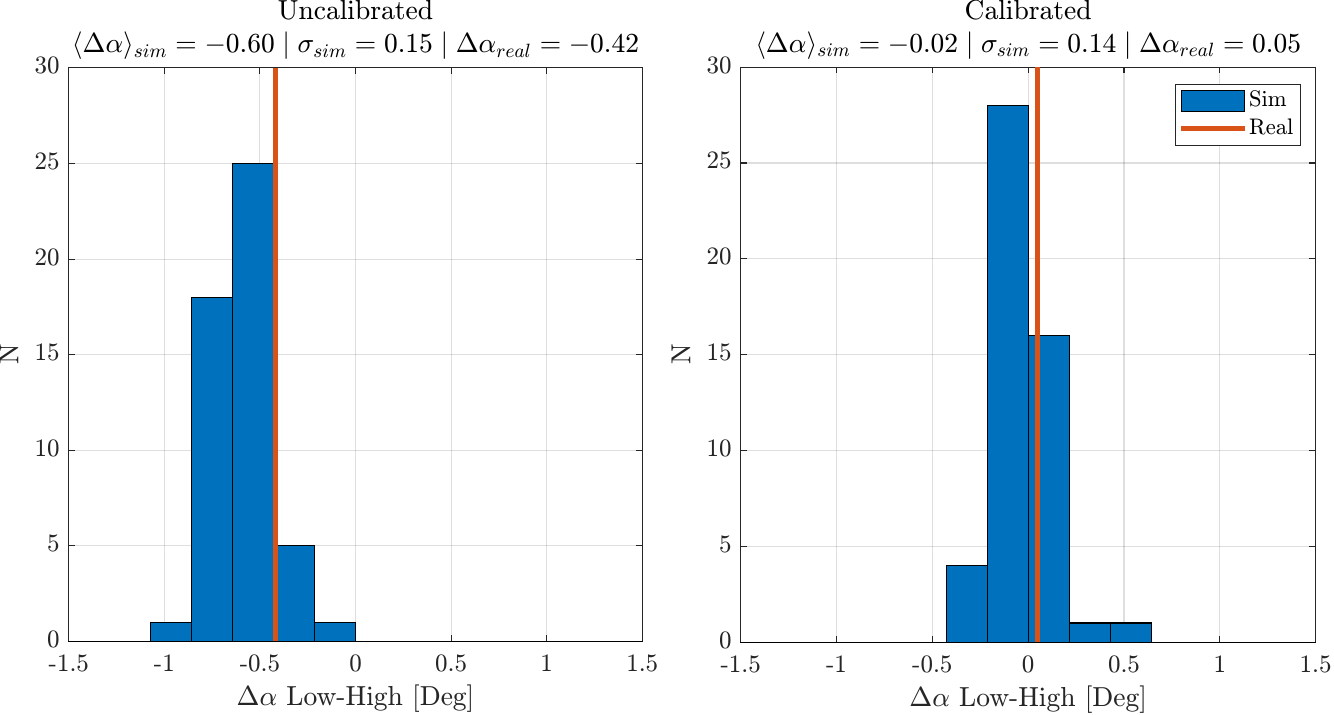}
    \end{adjustbox}
    \caption[$\Delta\alpha$ Legacy vs. Reanalysis Sims]{We compute the $\Delta \alpha$ quantity defined in Eq.~(\ref{eq:delta_alpha}) for the uncalibrated (\textbf{left}) and calibrated (\textbf{right)} approach, on both simulations (blue histograms) and real data (red line). The difference between the ``high''and ``low'' subsets goes away as expected when using RPS measured angles, which validates that these angles are a better representation of reality than idealized assumptions used in the uncalibrated approach.}
    \label{fig:dalphahists}
\end{figure*}

We now repeat the procedure on real data --- since we are looking only at differences, we remain blind to the real celestial angle.
As shown by the red lines in Fig.~\ref{fig:dalphahists}, the real data agree well with both approaches.
This shows that the relative distribution of angles measured using the RPS is a better representation of the reality than our ideal assumptions.
The fact that the difference correctly goes to zero in the real data for the calibrated approach reassures us that the angles are being properly accounted for in the analysis (i.e., no sign errors, etc.).

This procedure therefore lends confidence that the relative measurements are accurate and are being correctly taken into account in the analysis. 
It does not give any information on the absolute angles, from the sky or from a common mode systematic effect, which would cancel out in the difference.

\section{Constraining power for birefringence analysis}
\label{sec:cmb_uncertainty}
We derive the constraining power for birefringence analysis of our map data by analyzing a set of 50 simulations of the BICEP3 2-year dataset. 
We present the main sources of statistical and systematic uncertainty in Sec.~\ref{sec:B3}.
We conclude on the sensitivity of this dataset for birefringence searches and forecast future performance in Sec.~\ref{sec:forecast}.

\subsection{BICEP3 2-year dataset}
\label{sec:B3}
We split the on-sky sensitivity into the following contributions: instrument noise, lensing, and dust.
For each component of the sky model (CMB including lensing, noise, dust), we showcase two ways in which a given component contribution can be evaluated:

\begin{itemize}
    \item intrinsic EB power --- e.g., $E_{dust} \times B_{dust}$
    \item B modes from that component crossed with E modes from all other --- e.g., $B_{dust} \times E_{dust + noise + CMB}$, which we abbreviate to $B_{dust} \times E_{all}$ in the following. 
\end{itemize}

While the first term (intrinsic EB power) is expected to be small in most cases, the second term will be dominated by the chance correlations of bright CMB E modes crossed with B modes of each component.
This will give us a good representation of how each component contributes to the final statistical constraining power, which encompasses all of these terms. 

We estimate angles from 50 simulations in each case, and quote the standard deviation of the output angle distribution as the uncertainty coming from each component. 
For conciseness, instead of writing e.g. $\sigma_{\alpha,dust}$ to denote the uncertainty on the cosmic birefringence angle coming from dust, we abbreviate to $\sigma_{dust}$.
The first number we quote corresponds to the 14-bandpower case, and the number in parenthesis to the 9-bandpower case. 
In most cases, the constraining power is much better when using 14 bandpowers, as discussed in Sec.~\ref{sec:ell_choice}.

We first look into the relative contributions of instrument noise and lensing in sections~\ref{sec:noise} and~\ref{sec:lensing}.
We then explore in detail the impact of dust in the corresponding Sec.~\ref{sec:dust}. 
Finally, we use specialized simulations described in Sec.~\ref{sec:beams} to assess the impact of beam uncertainties and systematics.

\subsubsection{Instrument noise}
\label{sec:noise}

For this analysis, we use BICEP3 data taken in 2017 and 2018. 
We only use detectors for which we have a polarization angle measurement, corresponding to $\sim 90\%$ of detectors used in the standard CMB analysis.
This results in a map depth in polarization of $3.3 \mu$K.arcmin for the 2-year dataset.
As expected, the intrinsic contribution of noise to the EB signal is very low, with $\sigma_{noise} = 0.006\deg(0.008\deg)$ fitting over 14(9) bandpowers.
However, the contribution of noise when crossing with all signal types becomes our dominant sources of statistical uncertainty with $\sigma_{noise \times all} = 0.061\deg(0.078\deg)$.

\subsubsection{Lensing}
\label{sec:lensing}
The unlensed CMB signal contributes very little to the uncertainty budget, either individually ($\sigma_{unlensed} = 0.005\deg(0.009\deg)$) or when crossed with other components ($\sigma_{unlensed \times all} = 0.004\deg(0.004\deg)$). 
We would expect B modes to be zero in that case, but this contribution is due to residuals from E/B purification using matrix separation~\cite{BKVII}.
When including lensing B modes, we get $\sigma_{lensed \times all} = 0.035 \deg$ ($0.059 \deg$).
The increased statistical uncertainty is due to the increased variance of lensing B modes.
We did not run simulations containing only the lensed CMB signal (for both E and B modes), but the intrinsic EB contribution of such combination can expected to be very small, particularly for $\ell < 500 $~\cite{Naokawa2023}.

\subsubsection{Dust}
\label{sec:dust}
Dust emission is the principal contaminant to CMB polarization above 150GHz.
At lower frequencies, the signal is still present, but its amplitude is lower.
Many dust models predict that the dust EB signal is zero in the large ensemble average, without excluding localized fluctuations in a given sky patch. 
Recent work have also shown evidence of excess EB in large sky areas and discussed mechanisms to explain the generation of such signal~\cite{2021clark,cukierman2023,2023halal,2023vacher}.
Two mechanisms are at play here.
First, localized fluctuations increase the variance on the EB spectrum, and contribute to the statistical uncertainty. 
Additionally, chance correlations might bias the estimation of an isotropic angle if E and B modes were correlated enough to produce a locally non-zero EB signal.
Finally, one needs to consider how the interplay between CMB and dust can lead to increased statistical uncertainty and/or systematic bias when crossing CMB E modes with dust B modes, which can be expected to be bright, even at 95~GHz and in a clean sky patch. 

We simulate a subset of dust models and investigate their impact in the context of this analysis.
We follow the same approach as we did for lensing and instrument noise, separating the intrinsic EB contribution from the interplay between dust and other sky components.
We also pay attention to systematic contribution, i.e., biases on $\alpha$.

\paragraph{Choice of models}
We first recall that this work is limited to observations at 95~GHz in a $\sim600$ square degrees sky patch at high galactic latitudes, which naturally limits the impact of dust contamination.
In the BK patch, the dust signal can be accurately modeled by a Gaussian random field with a power spectrum parametrized by the dust amplitude $A_d$ and spectral index $\beta_d$, and no fiducial EB power.
We choose this model as the baseline for this analysis but also consider alternate dust models.
These alternate dust models are taken from the ones considered in previous \bk analysis~\cite{BKX,BKXIII} and in forecasting for CMB--Stage4~\cite{S4forecast}. 
More specifically, we consider only models for which the predicted dust amplitude in the \bk patch has been shown to be close to the measured one.
A summary of models considered here is shown in Table \ref{tab:dust_models} --- one can refer to Appendix E4 of \cite{BKXIII} for a complete description.

\begin{table}
\caption{Dust models considered in this analysis}
\begin{tabular}{ lc }
\hline
\hline
\multirow{2}{*}{\textbf{Dust model} } & \textbf{Best fit dust amplitude} $\mathbf{A_d [\mu K ^2]}$ \\
&\textbf{in the BK patch} \\ 
\hline

\hline
Gaussian (baseline) & 3.9 \\ 
MKD \cite{MKD} & 3.9 \\
Vansyngel \cite{Vansyngel} & 5.5\\ 
MHDv3 \cite{MHD1,MHD2} & 3.2 \\ 
\hline
\hline
\end{tabular}
\label{tab:dust_models}
\end{table}

\paragraph{Dust EB contribution}
We first explore how variance in the intrinsic EB dust signal contributes to the statistical uncertainty. 
We fit an angle to 50 simulations containing only Gaussian dust.
The statistical uncertainty due to dust EB only is extremely low, with $\sigma_{G.\,dust} = 0.0007\deg$, with no systematic bias as expected.

For alternate dust models, we cannot repeat the same procedure because these models do not consider statistical variations in dust, and therefore we only have one realization of each model.
The way these models affect the total statistical uncertainty is discussed in paragraph \textit{c} below.
Additionally, we compute dust-only EB spectrum of each model, and find that none of them is susceptible to bias the angle by more than $0.004^\circ$ ($0.009^\circ$). 
We emphasize again that these results are for single realizations and therefore do not take into account the uncertainty on each of these models.

Finally, we consider a toy model of maximally correlated dust. 
For each model, we take the EB bandpowers to be the geometrical mean of the EE and BB bandpowers:

\begin{equation}
\mathcal{C}_b^{EB} = \sqrt{\mathcal{C}_b^{EE} \times \mathcal{C}_b^{BB}}.
\end{equation}
Across all models, this yields to a maximum bias on the angle of $0.027 \deg$ ($0.058 \deg$) for 14 (9) bandpowers. 
This scenario is obviously unrealistic but has the merit to show that even if the dust E and B modes were highly correlated, the bias on the angle would still be very small for models with dust amplitudes similar to what is measured in the \bk patch.

\paragraph{CMB --- Dust chance correlations}

In the previous section, we only focused on the contribution of the intrinsic dust EB signal. 
Here we investigate the chance correlations of dust B modes with sky E modes (lensed CMB + noise).
We take the E modes from our standard simulations, and cross them with B modes from different dust models and then fit angles to these cross spectra. 
The uncertainty that we quote here comes from the variance in CMB realizations, but we still have a single dust realization.
For Gaussian dust, we find no bias and a variance of $0.007^\circ$ ($0.016^\circ$).
None of the alternate models results in a significant bias or increased variance compared to Gaussian dust, except for MHDv3 for which we find a small excess at the level of $0.016 \deg \pm 0.011 \deg$ ($0.037 \deg \pm 0.021 \deg$). 

\paragraph{BK dust map}

Finally, we use a dust-only map of the BICEP/\textit{Keck} patch, obtained by combining data at multiple frequencies and performing map-based component separation~\cite{BKXIX}.
We repeat the same analysis as for the dust models, taking into account the specific noise properties of this map.
The dust-only EB signal extracted from this map is consistent with simulations of EB-free Gaussian dust + noise, and we find no bias associated with it.
For the chance correlation between dust B modes and other component E modes, we find a small angle compatible with Gaussian dust.
If we assume maximal correlation between E and B in that map, the maximum bias on the angle is $0.02\deg$, similar to the dust models that we considered. 

\paragraph{Summary}
We have shown that:

\begin{itemize}
    \item The statistical uncertainty coming from Gaussian dust alone is negligible.
    \item When combining Gaussian dust with other sky components, we get $\sigma_{G.\,dust \times all} = 0.007^\circ$ ($\sigma_{G.\,dust \times all} = 0.016^\circ$) for 14 (9) bandpowers. We include that number in our uncertainty budget as the statistical uncertainty coming from dust in our patch.
    \item The alternate dust models and real data that we have considered do not affect the total statistical uncertainty by a significant amount compared to the baseline Gaussian model case.
    \item Our toy model of maximally correlated dust shows that dust is unlikely to significantly bias the angle estimate for any model with dust amplitude similar to that observed in the \bk patch.
    \item When considering only dust B modes crossed with CMB + noise E modes, one model shows a modest excess in EB leading to a positive value for $\alpha$. 
\end{itemize}

To be conservative, we therefore choose to assign an upper limit on systematic uncertainty coming from dust of $\sigma_{syst,dust} = 0.02 \deg$ ($\sigma_{syst,dust} = 0.04\deg$) for 14 (9) bandpowers.
We emphasize again that these results are valid only in the BICEP3 observing field and at 95~GHz.

\subsubsection{Instrumental systematics}
\label{sec:beams}

Beam effects are one of the dominant sources of instrumental systematics for CMB experiments.
In particular, for experiments using pair difference to reconstruct the polarized signal, beam mismatch between detectors in the same pair is a known source of instrumental temperature-to-polarization leakage.
For BICEP/\textit{Keck}, this is the dominant source of instrumental systematic effects \cite{BKXI,BKIII}. 
To mitigate such effects, we rely on high fidelity, far field beam measurements and deprojection techniques for the lowest order modes of the differential beam response~\cite{BKIII,BKIV,2016karkare,2020stgermainespie}.

However, undeprojected differential beam residuals are still a significant source of systematic error. We explore their impact on the EB signal in Sec.~\ref{sec:tpleakage}.
Additionally, because we are using higher bandpowers in this analysis compared to previous \bk publications, we need to pay particular attention to beam window function uncertainties at high multipoles, which we investigate in Sec.~\ref{sec:bls}.

\paragraph{Undeprojected beam residuals}
\label{sec:tpleakage}

Undeprojected beam residuals are responsible for temperature-to-polarization leakage.
In the mainline CMB analysis, we use high fidelity beam measurements to produce specialized simulations called beam sims, whose goal is to predict the impact of temperature-to-polarization leakage on the BB power spectrum.
We then cross this template for temperature-to-polarization leakage with our real CMB data and perform a full multi-component analysis to assess the bias on \textit{r} of that effect~\cite{BKXI,BKXIII}.

\begin{figure*}
    \centering
    \begin{adjustbox}{width={\textwidth},keepaspectratio}
\includegraphics[width = \textwidth]{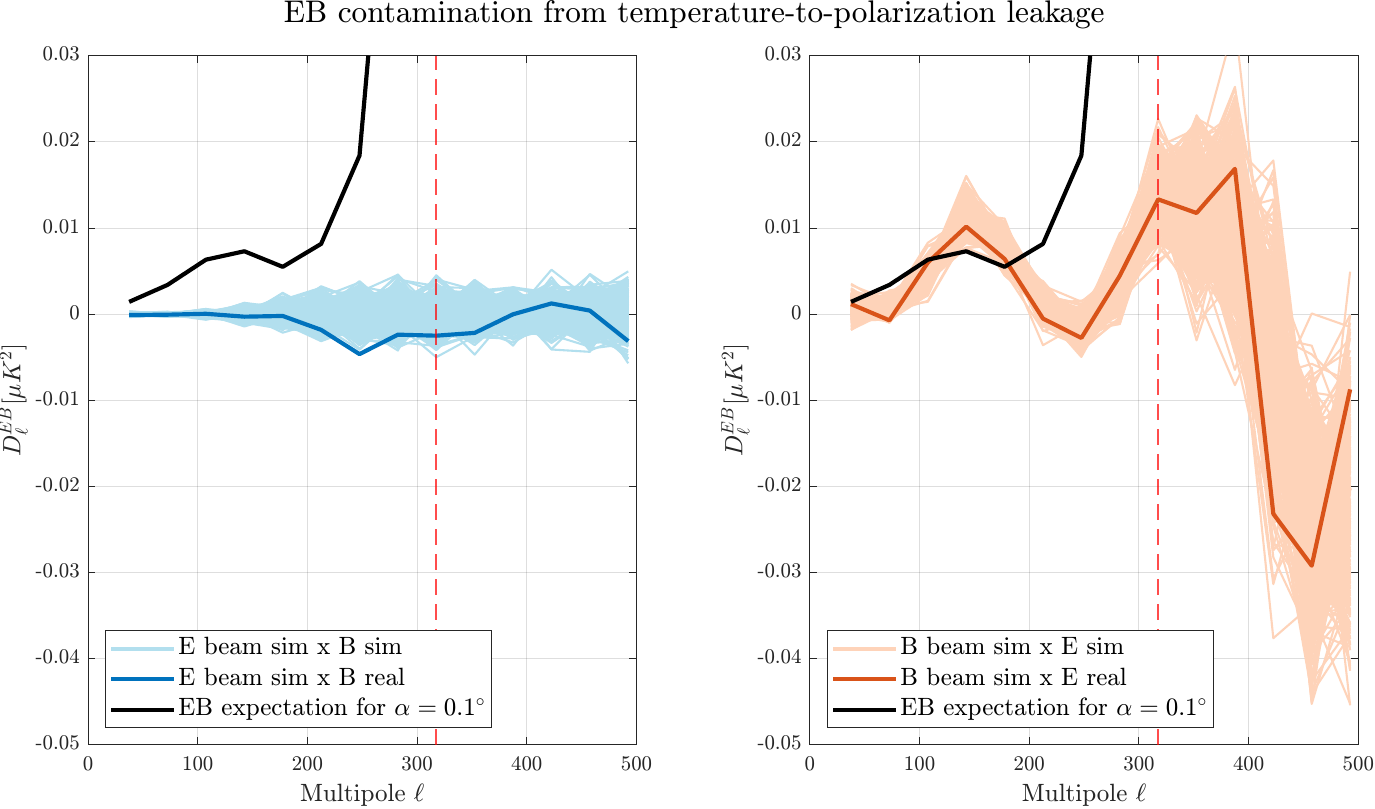}
    \end{adjustbox}
    \caption{EB contamination from temperature-to-polarization leakage. For each panel, the darker line is the cross with real data, and the lighter lines show the cross with 499 signal + noise + dust simulations. The black line is the expectation value for the EB spectra from isotropic polarization rotation for $\alpha = 0.1\deg$.
    The vertical dash line is the separation between the 9- and the 14-bandpower case.
    The \textbf{left} panel shows the contribution from E beam sim $\times$ B CMB, which is noise dominated and well below the equivalent $\alpha = 0.1\deg$. The \textbf{right} panel shows the contribution from B beam sim $\times$ E CMB, driven by TE correlations in the CMB and T-to-B leakage from the instrument.}
    \label{fig:tpleakage}
\end{figure*}

For the present analysis, we use the same framework and specialized beam simulations to predict the shape and amplitude of the EB signal contamination.
To do so, we again cross the temperature-to-polarization leakage template with our real CMB data and a set of simulations, but we look at the EB signal instead of the BB one.
In that context, there are two terms that can contribute to temperature-to-polarization leakage contamination of the EB signal, as shown in Fig.~\ref{fig:tpleakage}:
\begin{itemize}
    \item E modes sourced by beam T-to-P leakage crossed with the sky B modes;
    \item B modes sourced by beam T-to-P leakage crossed with the sky E modes.
\end{itemize}
We find the first term to be dominated by noise in the sky B modes, and to contribute only a small part to the EB signal contamination.
The second term, however, is more significant.
On one side of this cross spectrum, the temperature-to-polarization leakage mechanism leads to a small B modes signal, naturally correlated with the sky temperature signal which sources it. 
On the other side, CMB E modes are also correlated with CMB temperature. 
It ensues that this particular EB cross spectrum follows the real TE correlations in the CMB, which results in a non-negligible contamination.

We use our usual procedure to estimate the impact on the sky angle of this effect, by fitting angles to EB power spectra.
We find that the bias on the angle is $0.028\deg \pm 0.0047\deg$ ($0.071\deg \pm 0.0076\deg$) for 14 (9) bandpowers on real data, where the uncertainty comes from repeating the same procedure on simulations.
The bias is therefore clearly detected above zero, and is in excellent agreement with simulations as shown in Fig.~\ref{fig:tpleakage}.
The difference between the two cases is due to the specific shape of the signal that resembles the prediction for isotropic rotation at low $\ell$ but not at higher $\ell$, as shown in the right panel of Fig.~\ref{fig:tpleakage}.
To also take into account the small contribution of the other term (right panel of Fig.~\ref{fig:tpleakage}, we adopt a conservative additional systematic uncertainty of $0.03\deg$ ($0.08\deg$) for 14 (9) bandpowers.

\paragraph{Beam window function}
\label{sec:bls}

Another source of systematic effect is the uncertainty on the beam window functions ($B_\ell$) that we derive from far field beam mapping data.
We have evidence that such errors would scale with $\ell$, as it is much harder to get high confidence measurements at small angular resolution (high $\ell$).
The beam window functions are used, amongst other things, to compute bandpower window functions.
An error on $B_\ell$ would therefore result in a multiplicative error on the power spectrum.

We estimate that a 5\% error on beam window function that scales with $\ell$ (larger error at high $\ell$) would lead to a 10\% (5\%) error on the best fit angle for 14 (9) bandpowers. 
Because it is multiplicative, such effect is not significant as long as we are estimating angles whose expectation values are close to zero.
This led us to derotating the uncalibrated maps as described in Sec.~\ref{sec:uncal_vs_cal} to bring their expectation values to zero.

\subsection{Summary \& Perspectives}
\label{sec:forecast}

\subsubsection{Summary}

All dominant sources of uncertainties for cosmic birefringence constraints coming from BICEP3 sky maps are outlined in Table \ref{tab:cmb_uncertainties} for the 14-bandpower case.
We also quote our total statistical uncertainty estimated from full signal simulations containing all sky components.
This establishes the statistical constraining power of this dataset for cosmological birefringence at $\sigma_\alpha = 0.078 \deg$.
One can note that this number is within 10\% of adding the dominant sources of uncertainties in quadrature.
We note in particular that the dominant source of statistical uncertainty for this two-year dataset is instrument noise, followed by lensing variance.
The contribution of dust remains at negligible levels in our sky patch at 95~GHz, even for non-Gaussian dust models.
It is also interesting to note that the systematic error due to temperature-to-polarization leakage is a small but not negligible contribution to the total error budget.
In future analysis, if this effect continues to be reliably detected, one could use the measured leakage signal to debias the EB spectra prior to estimating the birefringence angle. 

\begin{table}
\caption{Summary of the on-sky sensitivity to the birefringence angle for the BICEP3 2-year dataset. Only the main contributions discussed in the previous paragraphs are listed here. The \textbf{Total} number however does include all contributions including ones that we have not specifically detailed, as they account for less than 10\% of the final sensitivity.}
\begin{tabular}{ lll }
\hline
\hline
\multicolumn{1}{c}{\textbf{Type}} & \multicolumn{1}{c}{\textbf{Source}} & \multicolumn{1}{c}{\textbf{$\sigma_{\alpha}$}} \\ 
\hline

\hline
\multirow{4}{*}{\begin{tabular}[c]{@{}l@{}}Statistical\end{tabular}} & Instrument noise                                                          & $0.061 \deg$              \\ 
                                                                                   & Lensed CMB                                                                & $0.035 \deg$              \\ 
                                                                                   & Gaussian dust                                                             & $0.007 \deg$            \\ 
                                                                                   & \textbf{Total}                                                                     & $0.078 \deg$              \\ \hline
\multirow{2}{*}{\begin{tabular}[c]{@{}l@{}}Systematic\end{tabular}}  & \begin{tabular}[c]{@{}l@{}}Beam T-to-P leakage\end{tabular}      & $0.03 \deg$               \\ 
                                                                                   & \begin{tabular}[c]{@{}l@{}}Non-Gaussian dust\end{tabular} & $0.02 \deg$               \\ 
\hline
\hline
\end{tabular}
\label{tab:cmb_uncertainties}
\end{table}

\subsubsection{Projections for cosmic birefringence searches}
To conclude on our sensitivity to cosmic birefringence, we note that contributions of the individual components to the variance on the sky rotation angle $\sigma_\alpha^2$ add linearly. 
The same is true of BB power, and we can therefore derive a scaling relationship between BB power and $\sigma_\alpha^2$.
We can use that scaling relationship to make projections of the sensitivity we could reach by adding more data, as well as delensing to reduce lensing variance.

This work only uses 2 years of observations, 2017 and 2018, and BICEP3 has since been continuously observing (we recall 
that data taken in 2016 cannot be used for this analysis).
By adding 5 more years of data (2019 -- 2023), one could significantly reduce the contribution from instrumental noise to the total BB power.
This could reduce the uncertainty coming from instrumental noise down to $\sigma_{noise} = 0.035 \deg$, making the measurement lensing-variance limited.
One could also reduce the contribution from lensing using delensing techniques, as demonstrated in~\cite{2021BKSPT}.
Achieving 50\% delensing would reduce the lensing contribution to $\sigma_{lensing} = 0.024 \deg$.
Estimates of final on-sky sensitivity from these two improvements are summarized in Table~\ref{tab:forecast}.

These results, combined with RPS performance improvement, are compared to previous constraints in Table~\ref{tab:bcconstraints}. 
This work therefore has the potential to yield the most precise and accurate constraints on cosmic birefringence to date.

\begin{table}
\caption{Projections of on-sky sensitivity for BICEP3 when including more data and delensing. We recall that contributions to the constraining power $\sigma_\alpha$ add in quadrature, and that not all contributions are listed here. The sensitivity for future analysis scenarios is given as an estimate.}
\begin{tabular}{ llll }
\hline
\hline
\multicolumn{1}{c}{\textbf{Signal}} & \multicolumn{1}{c}{\textbf{$\sigma_{noise}$}} & \multicolumn{1}{c}{\textbf{$\sigma_{lensing}$}} & \multicolumn{1}{c}{$\mathbf{\sigma_{tot}}$} \\ 
\hline

\hline
B3 2 years (this work)                & $0.061\deg$                                   & $0.035\deg$                                     & $\mathbf{0.078 \deg}$                                 \\
B3 7 years                            & $0.035\deg$                                   & $0.035\deg$                                     & $\mathbf{0.055 \deg}$                                 \\ 
B3 2 years + delensing                & $0.061\deg$                                   & $0.024\deg$                                     & $\mathbf{0.073 \deg}$                                 \\ 
B3 7 years + delensing                & $0.035\deg$                                   & $0.024\deg$                                     & $\mathbf{0.048\deg}$                                  \\ 
\hline
\hline
\end{tabular}
\label{tab:forecast}
\end{table}

\subsubsection{Transfer of polarization calibration}
The sensitivity per components reported in Table \ref{tab:cmb_uncertainties} can be split between contributions coming from the sky signal (lensed CMB, dust) and contributions coming from the instrument (instrument noise and beam systematics). 
While both are relevant when it comes to constraining cosmic birefringence, the former becomes unimportant in the context of polarization calibration.

How well a CMB instrument can be calibrated on celestial sources is ultimately limited by the prior determination of the source polarization properties.
For another CMB telescope observing the same sky patch as BICEP3, the sky signal (CMB, foregrounds) is obviously identical.
The measurement of this same polarized signal can therefore serve to transfer polarization information, up to the precision and accuracy that this sky patch has been measured by BICEP3.
This is conceptually exactly the same thing as establishing the polarization properties of a compact source such as a planet or a nebula to serve as calibration standard.

This work therefore has the potential to make the BICEP3 sky patch a calibration source for CMB experiments, {\color{black} as long as filtering effects, including the beam transfer function, are correctly taken into account.}
The current precision on that calibration source would be $0.061\deg$, corresponding to the contribution of instrumental noise only.
The accuracy is presently limited by the determination of the absolute polarization angle (dominated by a systematic uncertainty on the RPS side), and by beam {\color{black} systematics due to T-to-P leakage}, but these could easily be taken out in future analysis. 
Additionally, as outlined in Table \ref{tab:forecast}, when considering all of the currently existing BICEP3 data, the noise contribution and therefore the calibration precision can be forecasted to reach $0.035\deg$.
This would provide a polarization calibration source for CMB experiments with a precision far exceeding those of the most commonly used celestial source, the Crab nebula~\cite{aumont2010,aumont2020}.

\section{Impact on inflation searches}
\label{sec:inflation}

\subsection{Methodology}
\subsubsection{Motivation}
As described in more details in Sec.~\ref{sec:uncal_vs_cal}, in the mainline CMB analysis, we do not use individual measurements of detector polarization angles to make maps. 
Instead we use design angles (with an overall adjustment coming from CMB pointing), which introduces an apparent miscalibration of the instrument polarization angle. 
The biggest manifestation of that effect is the generation of EB/TB signals of instrumental origin. 
It also generates a small BB signal, which is a potential source of systematic contamination for constraining $r$.

This effect is corrected for by fitting an overall polarization angle to the maps, that nulls the EB/TB spectra. 
The angle obtained from this fit can then be applied to rotate the BB signal and cancel out the effect of polarization rotation.
This process is commonly referred to as self-calibration, EB nulling, or calibration off the E modes. 
While it does correctly calibrate the overall polarization angle of the instrument, it does not take into account detector-to-detector variation in polarization angles and leaves some level of residual BB signal.

As we have shown in Sec.~\ref{sec:meas_results}, there are clear trends in detector polarization angles that cannot be corrected for by an overall rotation, with both tile-to-tile and per-tile variations.
Quite importantly, these patterns are not random but can be mapped to physical features of the focal plane, in particular the tile-to-tile variation which we have related to physical clocking of the detector tiles.
The level of residual BB power coming from such coherent variations has never been simulated.
Previous work by the \bk collaboration assumed random distribution of polarization angles on the focal plane \cite{BKIII}, and most current CMB experiments generally only address the challenge of calibrating the overall polarization angle ~\cite{2021abitbol,2021verges,2023jost}.

\subsubsection{Simulations \& Metric}
\paragraph{Simulations}
We use the simulations previously described in Sec.~\ref{sec:cmb_sims}, but looking at the BB power spectra instead of EB.
In particular, we want to compare how the different approaches perform --- uncalibrated derotated vs. calibrated. 
This is the difference between correcting for the overall angle vs. using measurements for each pair individually.
We also simulate a case where only tile-to-tile variations are included --- all pairs in a given tile are assumed to have the same polarization angle, given by the median over the tile. 
For that case, we process the data using the uncalibrated derotated approach: each tile has its own angle in TOD generation, then ideal angles are assumed in map making, and the spectra are derotated assuming a single polarization angle for the entire instrument.
Finally, we produce an extra set of simulations in which we use measured angles in TOD generation, and in map making we use the same measured angles but with an added random Gaussian error term of mean $\mu = 0.1\deg$ and standard deviation $\sigma = 0.05\deg$. 
This is an attempt at modeling a more realistic situation in which true detector angles are used in TOD generation, but for map making one would use angles that have been measured with some statistical uncertainty and systematic error.

\paragraph{Metric}
To compare the amount of residual BB power in each case, we use a quadratic estimator $\rho$ previously defined and used in \cite{2013aiken,BKXI}.
This is a single number that quantifies the amount of BB power compared to a nominal $r = 1$ power spectrum, for a given noise level and weighting scheme. 
In this case, we use the noise and weighting corresponding to the BICEP3 dataset, as detailed in e.g., \cite{2020StGermaine}.

\subsection{Results}
Results are summarized in Table \ref{table:r_bias}.
For reference, we also include the ``Uncalibrated'' case, corresponding to residual BB power prior to any calibration or derotation.
As expected, this case would lead to a large amount of residual BB power, close to the sensitivity of current CMB experiments, which underscores the importance of the global angle calibration procedure for inflation searches.
The ``Calibrated'' case also leads to an unsurprising result, with the residual BB power being completely eliminated.

The ``Uncalibrated derotated'' case is more interesting. 
The value of $\rho = 4 \pm 3 \times 10^{-5}$ corresponds the amount of residual BB power left after the global polarization angle has been taken out. 
This value remains very small, even compared to a sensitivity of $\sigma(r) = 5 \times 10^{-4}$ for, e.g., CMB-S4~\cite{cmbs42016}, and one can expect that the bias on $r$ would be even smaller.
Still, care should be taken in the future to make sure that such effects are not likely to bias inflation searches.
In particular, BICEP3 polarization angles have a relatively tight distribution with a standard deviation of $\sim 1 \deg$ for the detector-to-detector variation, and one would expect the residual power to grow as the detector-to-detector variation gets larger.

It is also interesting to note that when considering only tile-to-tile variations, we get a very similar value with $\rho = 4 \pm 1 \times 10^{-5}$.
This seems to indicate that the mean amount of residual BB power is driven by tile-to-tile variations, with the standard deviation driven by per-tile scatter.
This might prove insightful as we have shown in this paper that these relative tile clocking angles can be extracted from CMB data. 
{\color{black}In principle, these angles could be used to reduce residual BB power, but in practice for BICEP3, these angles would be noisier than those extracted from RPS data.}

Finally, the case ``Calibrated with error'' leads to a higher $\rho$ value than the ``Uncalibrated derotated'' procedure, which suggests that this latter method performs better in the case where angles are not precisely measured.
For current and future experiments, as long as the measurement uncertainty remains smaller than the detector-to-detector variation, a combination of calibration to address detector-to-detector variations + derotation to address the global offset angle is expected to be the better solution in the context of inflation searches, as shown by the case ``Calibrated with error + derotated''.

\begin{table}
\caption{$\rho$ estimates for various configurations, showing impact of residual B-modes power after mitigation by calibration and/or analysis (derotation).}
\begin{tabular}{lc}
\hline
\hline
\textbf{Case}          & \textbf{$\rho \left(10^{-5}\right)$} \\ 
\hline

\hline
Uncalibrated           & $510 \pm 64$                   \\ 
Calibrated             & $0 \pm 2$                      \\ 
Uncalibrated derotated --- all detectors & $4 \pm 3$                      \\ 
Uncalibrated derotated --- tile clocking only & $4 \pm 1$                      \\ 
Calibrated with error & $8 \pm 3$                      \\ 
Calibrated with error + derotated & $0 \pm 2$                      \\ \hline
\hline
\end{tabular}
\label{table:r_bias}
\end{table}

\section{Conclusion}
\label{sec:conclusion}

We have performed high precision measurements of BICEP3 polarization angles using a custom-made calibrator.
The internal consistency of our dataset is excellent, with a per-pair repeatability of $0.02\deg$.
We have performed several cross-checks with real data from CMB observations.
In particular, we have shown that tile-to-tile offsets measured using the RPS are in excellent agreement with CMB-derived detector pointings.
We also demonstrated that the relative distribution of angles measured using the RPS is a better representation of reality than an ideal assumptions about detector angles.

After a thorough review of measurement uncertainties, we are left with a systematic uncertainty at the level of $0.3\deg$, due to azimuth-dependent effects of the RPS/receiver alignment.
We have established that this systematic is sourced by a diffraction effect specific to this RPS assembly and components, and is not a fundamental limitation of our measurement method. 
In future work, we will therefore improve electromagnetic shielding and baffling of the RPS enclosure and shroud to mitigate it.
We will repeat tests on the improved apparatus to confirm that the effect can be reliably mitigated for future campaigns.
We have also installed a high precision optical encoder on the RPS. 
This will help mitigate the effect of the rotation stage backlash, which at the level of $0.06\deg$, is currently our second biggest source of measurement systematics.
With these two effects addressed, our calibration technique will be capable of reaching a measurement accuracy of $\mathcal{O}(0.05\deg)$.

We quantified the contributions of various components to BICEP3 on-sky sensitivity to cosmic birefringence.
We showed that the dominant source of uncertainty is currently instrument noise and established the on-sky sensitivity for the BICEP3 2-year dataset at $\sigma_\alpha = 0.078\deg$.
Adding all of the existing BICEP3 data through the 2023 observing season would decrease the contribution of instrumental noise by almost a factor two and make the measurement lensing limited, with an estimated sensitivity to the birefringence angle of $\sigma_\alpha = 0.055\deg$. 
Additionally, once systematic uncertainties have been cleared out, the BICEP3 sky patch can serve as a polarization calibration source for CMB experiments, which a precision forecasted to reach $0.035\deg$ for the 7-year dataset.

Finally, we showed that at the level that they are currently controlled, detector-to-detector polarization angle variations are unlikely to be a significant source of systematic error for B-mode measurements.
Measurements of polarization angles can still be reliably used to calibrate the effect of detector-to-detector scatter, as long as the uncertainty on the measurement is smaller that the detector scatter itself.

\section*{Acknowledgments}
The \bk projects have been made possible through a series of grants from the National Science Foundation most recently including 2220444-2220448, 2216223, 1836010, and 1726917.
The development of antenna-coupled detector technology was supported by the JPL Research and Technology Development Fund and by NASA Grants 06-ARPA206-0040, 10-SAT10-0017, 12-SAT12-0031, 14-SAT14-0009, \& 16-SAT-16-0002. 
The development and testing of focal planes
was supported by the Gordon and Betty Moore Foundation at Caltech. 
Readout electronics were supported by a Canada Foundation for Innovation grant to UBC. 
Support for quasi-optical filtering was provided by UK STFC grant ST/N000706/1. 
The computations in this paper were run on the Odyssey/Cannon cluster supported by the FAS Science Division Research Computing Group at Harvard University. 

We thank the staff of the U.S. Antarctic Program and the South Pole Station who have enabled this research. 
We thank Hans Boenish and Sam Harrison, the BICEP3 winterovers for the years 2017 and 2018, as well as Karsten Look (2022 winterover) for his help in the 2022 calibration campaign.
We also thank Marion Dierickx for her significant contributions during her tenure as the BICEP Operations Manager.

We thank Susan Clark, Brandon Hensley, and Léo Vacher for useful discussions about dust models and dust contamination for birefringence searches.
We thank Johannes Eskilt for his careful reading of our previous work, which has helped us address inaccuracies in our literature review and clear up inconsistencies in polarization orientation definition.
Lastly, we thank Bill Holzapfel for his insight with investigating instrumental systematics.

\newpage

\appendix
\section{Combining EB and TB angle estimators}
\label{sec:appendix_ebtb}

We start by creating linear estimators of the form
\begin{equation}
\begin{split}
&\hat{\alpha}_b^{XY} = \frac{d\alpha}{d\rCb^{XY}}\oCb^{XY}.\\
\end{split}
\end{equation}
In the small angle limit, the expectation values for rotated EB and TB become:

\begin{equation}
\begin{split}
&\rCb^{TB} = 2\alpha \oCb^{TE}\\
&\rCb^{EB} = 2\alpha \oCb^{EE},\\
\end{split}
\label{eq:alphamodelsimple}
\end{equation}
Where we assume that $\oCb^{BB} << \oCb^{EE}$, which we can calculate by taking the derivatives of Eq.~(\ref{eq:alphamodelsimple}):
    
\begin{equation}
\begin{split}
&\frac{d\rCb^{TB}}{d\alpha} = 2\oCb^{TE} \\
&\\
&\frac{d\rCb^{EB}}{d\alpha} = 2\oCb^{EE}. \\
\end{split}
\end{equation}
We then divide our observed spectra by the derivatives to get the estimators

\begin{equation}
    \begin{split}
    &\hat{\alpha}_b^{TB} = \frac{d\alpha}{d\rCb^{TB}}\oCb^{TB} = \frac{\oCb^{TB}}{2\oCb^{TE}}\\
    &\\
    &\hat{\alpha}_b^{EB} = \frac{d\alpha}{d\rCb^{EB}}\oCb^{EB} = \frac{\oCb^{EB}}{2\oCb^{EE}}.\\
    \end{split}
\end{equation}

We can compare the performance of the TB-only, EB-only, and EB+TB estimators via the variance/covariance of $\hat{\alpha}$ for EB, TB and between EB and TB. The covariance of two spectra can be rewritten as:

\begin{equation}
\cov \left(\Cb^{AB},\,\Cb^{CD}\right) = \frac{1}{k}\left(\Cb^{AC}\Cb^{BD}+\Cb^{AD}\Cb^{BC}\right),
\end{equation}
where $k$ is the number of modes averaged per band.
First, for EB only:

\begin{equation}
\begin{split}
\cov\left(\hat{\alpha}_b^{EB},\,\hat{\alpha}_b^{EB}\right) &=\frac{\cov\left(\oCb^{EB},\,\oCb^{EB}\right)}{4\oCb^{EE}{}^2}\\
&= \frac{\oCb^{EE}\oCb^{BB}+\oCb^{EB}{}^2}{4k\,\oCb^{EE}{}^2}.\\
\end{split}
\label{eq:ebcov1}
\end{equation}
Assuming that the only source of EB signal is from rotation, the EB signal is negligibly small compared to EE, and we are left with

\begin{equation}
\begin{split}
\cov\left(\hat{\alpha}_b^{EB},\,\hat{\alpha}_b^{EB}\right)
&= \frac{\oCb^{BB}}{4k\,\oCb^{EE}},\\
\end{split}
\end{equation}
and for TB only, we can make the same assumption that the TB term also contributes negligibly:

\begin{equation}
\begin{split}
\cov\left(\hat{\alpha}_b^{TB},\,\hat{\alpha}_b^{TB}\right)&= \frac{\oCb^{TT}\oCb^{BB}+\tozero{\oCb^{TB}}^2}{4k\,\oCb^{TE}{}^2}\\
&\\
&= \frac{\oCb^{TT}\oCb^{BB}}{4k\,\oCb^{TE}{}^2}.\\
\end{split}
\end{equation}
    
Both TE and EE are large compared to TB and EB, allowing us to make the same argument as above for EB+TB:
\begin{equation}
\begin{split}
\cov\left(\hat{\alpha}_b^{TB},\,\hat{\alpha}_b^{EB}\right)
&= \frac{\oCb^{TE}\oCb^{BB}+\tozero{\oCb^{TB}\oCb^{EB}}}{4k\,\oCb^{TE}\oCb^{EE}}\\
&\\
&= \frac{\tozero{\oCb^{TE}}\oCb^{BB}}{4k\,\tozero{\oCb^{TE}}\oCb^{EE}}\\
&\\
&= \frac{\oCb^{BB}}{4k\,\oCb^{EE}},
\end{split}
\end{equation}
and we can see that the covariance is the same as for the EB-only case.

\bibliography{references}

\end{document}